\documentclass[aps,prl,reprint,twocolumn,superscriptaddress,floatfix,nofootinbib,longbibliography]{revtex4-2}
\usepackage{graphicx,amsmath,amsfonts,amssymb,amsthm}
\usepackage{dsfont,upgreek,physics}
\usepackage{mathrsfs}
\usepackage{widetable}
\usepackage{mathtools}
\usepackage{bbold}
\usepackage{float}
\usepackage{placeins}

\usepackage[bookmarks=true,hidelinks]{hyperref}
\usepackage[dvipsnames]{xcolor}

\AtBeginDocument{\let\oldcontentsline\contentsline}
\newcommand{\notoccontentsline}[4]{\oldcontentsline{}{}{}{}}
\newcommand{\droptocpage}{\addtocontents{toc}{\let\protect\contentsline\protect\notoccontentsline}}
\newcommand{\incltocpage}{\addtocontents{toc}{\let\protect\contentsline\protect\oldcontentsline}}

\begin{document}
\title{Coherent meson--baryon conversion in a cold-atom non-Abelian lattice gauge simulator}

\author{Tian-Yi Wang}
\thanks{T.-Y.W.~and W.-Y.Z.~contributed equally to this work.}
\affiliation{Hefei National Research Center for Physical Sciences at the Microscale and School of Physical Sciences, University of Science and Technology of China, Hefei 230026, China}
\affiliation{Hefei National Laboratory, University of Science and Technology of China, Hefei 230088, China}

\author{Wei-Yong Zhang}
\thanks{T.-Y.W.~and W.-Y.Z.~contributed equally to this work.}
\affiliation{Hefei National Research Center for Physical Sciences at the Microscale and School of Physical Sciences, University of Science and Technology of China, Hefei 230026, China}
\affiliation{Hefei National Laboratory, University of Science and Technology of China, Hefei 230088, China}

\author{Tie-Jin Zhong}
\affiliation{Hefei National Research Center for Physical Sciences at the Microscale and School of Physical Sciences, University of Science and Technology of China, Hefei 230026, China}
\affiliation{Hefei National Laboratory, University of Science and Technology of China, Hefei 230088, China}

\author{Ming-Gen He}
\affiliation{Hefei National Research Center for Physical Sciences at the Microscale and School of Physical Sciences, University of Science and Technology of China, Hefei 230026, China}
\affiliation{Hefei National Laboratory, University of Science and Technology of China, Hefei 230088, China}

\author{Zhao-Yu Zhou}
\affiliation{Hefei National Research Center for Physical Sciences at the Microscale and School of Physical Sciences, University of Science and Technology of China, Hefei 230026, China}
\affiliation{Hefei National Laboratory, University of Science and Technology of China, Hefei 230088, China}

\author{Hui Sun}
\affiliation{Hefei National Laboratory, University of Science and Technology of China, Hefei 230088, China}

\author{Zhen-Sheng Yuan}
\affiliation{Hefei National Research Center for Physical Sciences at the Microscale and School of Physical Sciences, University of Science and Technology of China, Hefei 230026, China}
\affiliation{Hefei National Laboratory, University of Science and Technology of China, Hefei 230088, China}
\affiliation{New Cornerstone Science Laboratory, University of Science and Technology of China, Hefei 230026, China}

\author{Jian-Wei Pan}
\affiliation{Hefei National Research Center for Physical Sciences at the Microscale and School of Physical Sciences, University of Science and Technology of China, Hefei 230026, China}
\affiliation{Hefei National Laboratory, University of Science and Technology of China, Hefei 230088, China}

\date{\today}

\begingroup
\hypersetup{allcolors=black}
\begin{abstract}

Quantum chromodynamics (QCD), the non-Abelian gauge theory of quarks and gluons, provides a microscopic description of strongly interacting matter, including hadron structure and dynamics \cite{wilson1974confinement,busza2018heavy}.
Real-time dynamics and finite-density properties of these systems remain challenging for conventional Monte Carlo methods because of sign problems \cite{de2010simulating,banuls2020simulating}.
Quantum simulation provides a complementary approach, although circuit depth and noise limit digital implementations and non-Abelian gauge--matter interactions remain difficult to engineer in analog systems \cite{ciavarella2025string,ilcic2026observation,halimeh2024spin}.
Here we realize a non-Abelian lattice gauge model with dynamical matter in an analog cold-atom simulator and directly observe its hadronic dynamics using site- and spin-resolved quantum gas microscopy. 
This realization is enabled by harnessing two hyperfine states and site-occupation configurations of \(^{87}\mathrm{Rb}\) atoms to jointly encode both matter and gauge fields. We tailor resonant tunneling in a spin-dependent optical superlattice to implement a gauge sector of a truncated one-dimensional \(\mathrm{SU}(2)\) quantum link model with independently tunable mass and gauge coupling.
Leveraging this control, we track the conversion of an initially mesonic state into baryonic or vacuum-like configurations through slow mass ramps.
Near resonance, we observe coherent oscillations between mesons and baryon–antibaryon pairs and characterize the conversion efficiency as a function of mass, gauge coupling and net baryon density.
Our realization establishes an atomic platform for non-Abelian gauge dynamics, opening a route to microscopic studies of nonequilibrium dynamics in strongly interacting matter.

\end{abstract}

\maketitle
\droptocpage

The strong interaction binds quarks into hadrons and determines the structure and dynamics of nuclear matter \cite{wilson1974confinement,busza2018heavy}.
Its microscopic description, quantum chromodynamics (QCD), is a non-Abelian gauge theory in which the color-carrying gluons couple to quarks and interact with one another \cite{gross1973ultraviolet,politzer1973reliable,bauer2023quantum}.
At hadronic energy scales, strong coupling between quarks and gluons requires a nonperturbative treatment of hadron structure and dynamics \cite{wilson1974confinement,andronic2018decoding}.
Lattice gauge theory (LGT) formulates QCD on a spacetime lattice \cite{wilson1974confinement,kogut1975hamiltonian}, enabling precise nonperturbative calculations of the hadron spectrum in Euclidean time \cite{durr2008ab}.
By contrast, real-time evolution involves oscillatory path-integral weights, whereas a finite baryon chemical potential generally makes the QCD fermion determinant complex \cite{de2010simulating}.
The resulting sign problems limit conventional Monte Carlo studies of real-time dynamics and finite-density matter \cite{troyer2005computational,banuls2020simulating}.

Quantum simulators approach these problems by evolving a controlled quantum system under the Hamiltonian of the target gauge theory \cite{feynman1982simulating,bauer2023quantum}.
Digital processors have demonstrated hadron spectroscopy, finite-density state preparation and early-time dynamics in non-Abelian models \cite{atas2021hadrons,than2025phase,ciavarella2025string,ilcic2026observation}, including string breaking in a pure-gauge theory \cite{john2026non}.
Hardware-adapted qudit encodings reduce the circuit resources required for gauge-theory simulations \cite{gonzalez2022hardware,zache2023fermion,meth2025simulating}.
However, longer Trotterized evolution at fixed accuracy generally requires deeper circuits, increasing sensitivity to gate errors and decoherence \cite{jordan2012quantum,ciavarella2025string,ilcic2026observation}.
Analog simulators instead implement the gauge Hamiltonian through engineered many-body interactions, enabling continuous evolution without gate-by-gate operations \cite{banerjee2012dynamical,schweizer2019floquet,mil2020scalable,halimeh2025cold}.
This approach has enabled microscopic cold-atom studies of confinement and string breaking in extended Abelian LGTs \cite{yang2020observation,zhang2025observation,liu2025string,gonzalez2025observation}.
Non-Abelian gauge dynamics requires matter--gauge couplings that satisfy local constraints generated by noncommuting operators \cite{tagliacozzo2013simulation,halimeh2022protection,halimeh2024spin}.
Realizing these couplings in a cold-atom system with dynamical matter and microscopic readout remains an outstanding experimental challenge, despite several concrete theoretical proposals \cite{banerjee2013atomic,dasgupta2022cold,surace2024scalable,halimeh2024spin,depaciani2026fermionic}.

\begin{figure*}[t]
    \centering
    \includegraphics[width=\textwidth]{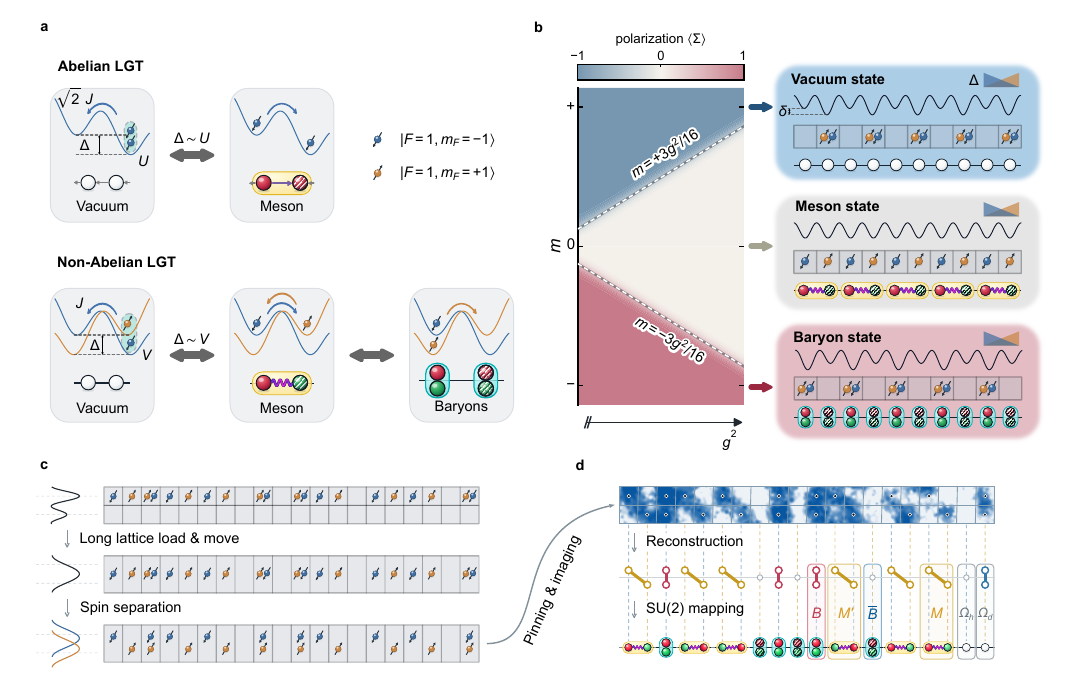}
    \caption{\textbf{A cold-atom non-Abelian lattice gauge simulator with microscopic readout.}
    \textbf{a,} Abelian vacuum--meson coupling (upper) is extended to include baryon--antibaryon configurations in the non-Abelian implementation (lower).
    A magnetic-field gradient produces opposite nearest-neighbor offsets $\mp\Delta$ for the two atomic spin states $\alpha$ and $\beta$ (blue and orange).
    Near $\Delta=V$, where $V$ is the interspecies interaction, first-order tunneling with amplitude $J$ couples a nearest-neighbor pair $\ket{\alpha,\beta}$ to the doublon states $\ket{\alpha \beta, 0}$ and $\ket{0, \alpha \beta}$, in which one $\alpha$ and one $\beta$ atom share a site.
    In this encoding, these configurations represent a meson, a baryon--antibaryon pair, and the vacuum.
    \textbf{b,} Schematic hadronic regimes in the mass--coupling plane at zero net baryon number.
    Dashed lines mark the zero-tunneling boundaries $m=+3g^2/16$ and $m=-3g^2/16$, separating the mesonic regime from the vacuum and baryonic regimes, respectively.
    Colors show the mean staggered atomic polarization; cartoons give the associated atomic and hadronic configurations.
    \textbf{c,} Transverse Stern--Gerlach separation maps the two spin components onto neighboring readout rows before pinning and fluorescence imaging.
    This spin-to-position mapping resolves holes, single $\alpha$ or $\beta$ atoms and $\alpha \beta$ doublons, which would otherwise be lost due to light-assisted collisions during imaging.
    \textbf{d,} Selected 20-site fluorescence snapshot (top), reconstructed atomic occupations (middle) and corresponding hadronic configuration (bottom).
    Hadronic identities depend on both the measured occupation and the matter sublattice through the staggered particle--hole encoding.
    $M$ and $M'$ label mesons on the two bond sublattices, $B$ and $\bar B$ baryons and antibaryons, and $\Omega_h$ and $\Omega_d$ vacuum sites encoded by holes and doublons.}
    \label{fig:encoding}
\end{figure*}

Here we report, to our knowledge, the first cold-atom realization of a non-Abelian lattice gauge theory with dynamical matter, using ultracold $^{87}$Rb atoms in a spin-dependent optical superlattice \cite{yang2017spin}.
Opposite energy gradients for the two hyperfine states combine with interspecies interactions to selectively bring first-order tunneling processes into resonance.
Within the resonant subspace, the atomic Hamiltonian maps onto a gauge sector of a truncated one-dimensional SU(2) quantum link model \cite{chandrasekharan1997quantum,wiese2022quantum}.
The staggered potential and interaction--gradient detuning independently control the mass and gauge coupling, while quantum gas microscopy reconstructs hadronic populations and color-electric flux \cite{bakr2009quantum,sherson2010single,boll2016spin}.
Slow mass ramps distinguish mesonic, baryonic and vacuum regimes and map the conversion boundaries between them.
When mesonic and baryon--antibaryon configurations are nearly degenerate in energy \cite{kasper2020jaynes}, time-resolved measurements reveal repeated baryon--antibaryon production and meson regeneration \cite{calajo2025scarring}.
We then quantify conversion efficiency along mass and gauge-coupling scans and investigate its density dependence in ensembles with finite net baryon density.

\section{\label{sec:mapping}Implementing a non-Abelian lattice gauge model with atoms}

In our earlier Abelian lattice gauge simulator, a magnetic-field-induced tilt compensated the on-site interaction of single-component bosons, enabling resonant tunneling between $\ket{2,0}$ and $\ket{1,1}$ and hence vacuum--meson conversion \cite{zhang2025observation} (Fig.~\ref{fig:encoding}a, upper).
A meson is a particle--antiparticle composite with zero net baryon number (a measure of the particle--antiparticle imbalance).
Unlike our earlier Abelian realization, the non-Abelian extension supports color-singlet baryons and antibaryons made entirely of quarks or antiquarks, carrying net baryon numbers $+1$ and $-1$, respectively \cite{atas2021hadrons,banerjee2013atomic,raychowdhury2020loop}.
To encode these hadrons and their interconversion, we use ultracold $^{87}$Rb atoms in the hyperfine states $\alpha=\ket{F=1,m_F=-1}$ and $\beta=\ket{F=1,m_F=+1}$ in a spin-dependent optical superlattice \cite{yang2017spin}.
Their opposite magnetic moments produce opposite nearest-neighbor energy offsets $\mp\Delta$ under the same magnetic-field gradient.
Near $\Delta=V$, where $V$ is the interspecies interaction, first-order tunneling couples $\ket{\alpha,\beta}$ to both $\ket{\alpha\beta,0}$ and $\ket{0,\alpha\beta}$, each with matrix element $-J$ (Fig.~\ref{fig:encoding}a, lower).
On a particle--antiparticle bond, these states encode a meson, a baryon--antibaryon pair and the vacuum, respectively, making both meson creation and meson--baryon conversion accessible.
Connecting these building blocks along the lattice, with consistent states on shared links, implements a gauge sector of a one-dimensional SU(2) quantum link model.

The SU(2) quantum link model is described by the Hamiltonian \cite{kogut1975hamiltonian,atas2021hadrons}
\begin{equation}
\begin{split}
\hat H_{\rm LGT}={}&-\kappa\sum_{\ell}\sum_{c,c'\in\{\mathrm r,\mathrm g\}}
\bigl(\hat\psi_{\ell c}^{\dagger}\hat U_{\ell}^{cc'}
\hat\psi_{\ell+1,c'}+\mathrm{h.c.}\bigr)\\
&+m\sum_{\ell,c}(-1)^{\ell}\hat n_{\ell c}
+\frac{g^2}{2}\sum_{\ell}\sum_{a=1}^{3}
\bigl(\hat E_{\ell}^{a}\bigr)^2 ,
\end{split}
\label{eq:main_target}
\end{equation}
\noindent where $\hat\psi_{\ell c}$ is the fermionic matter-field operator at site $\ell$ with color $c\in\{\mathrm r,\mathrm g\}$, and $\hat n_{\ell c}=\hat\psi_{\ell c}^{\dagger}\hat\psi_{\ell c}$ is the corresponding number operator.
On each link, the SU(2) gauge field is described by the parallel-transporter matrix elements $\hat U_{\ell}^{cc'}$ and the color-electric-field operators $\hat E_{\ell}^{a}$.
The noncommuting Gauss operators $\hat G_{\ell}^{a}=\tfrac12\sum_{c,c'}\hat\psi_{\ell c}^{\dagger}\sigma^{a}_{cc'}\hat\psi_{\ell c'}+\hat R_{\ell-1}^{a}+\hat L_{\ell}^{a}$ generate local SU(2) transformations and commute with $\hat H_{\rm LGT}$.
Here $\sigma^a$ are Pauli matrices, and $\hat L_\ell^a$ and $\hat R_\ell^a$ act at the left and right ends of link $(\ell,\ell+1)$, with $\hat{\mathbf L}_\ell^2=\hat{\mathbf R}_\ell^2=\hat E_\ell^2$.
In this work, we truncate the gauge links at $j_{\max}=\tfrac12$, retaining the $j=0,\tfrac12$ representations \cite{zohar2015formulation}, and work in the invariant nonzero Gauss-charge sector $\hat{\mathbf G}_{\ell}^{2}=2$, $\hat G_{\ell}^{z}=1$.
This sector describes SU(2) gauge dynamics in a uniform background of static adjoint (spin-1) color charges~\cite{banerjee2018quantum}.
In the corresponding bare vacuum, dynamical matter excitations are absent and every gauge link occupies the $j=\tfrac12$ flux state.
The local gauge constraints restrict hadrons in this sector to compact bound states.
Mesons consist of neighboring quark--antiquark pairs joined by a $j=0$ link, whereas baryons and antibaryons are on-site two-quark and two-antiquark color singlets.
This constrained basis is directly encoded by the atomic configurations introduced above (Fig.~\ref{fig:encoding}a, lower).
Matching the projected matrix elements and energies gives $\kappa=\sqrt2J$, $m=\delta/2$ and $g^2=8(V-\Delta)/3$, where $\delta$ is the staggered superlattice offset.
The projected atomic Hamiltonian reproduces the allowed transitions and matrix elements of the encoded gauge sector.
Comparisons of dilute-hadron dynamics and strong-coupling spectra reveal common features in this sector and the source-free Gauss-singlet SU(2) sector, distinct from those of a matched two-flavor U(1) model \hyperref[sm:lgt:sector]{(Supplementary Information)}.
The atomic filling sets the conserved net baryon number, providing access to hadron dynamics at finite net baryon density \cite{atas2021hadrons,banerjee2013atomic}.

At zero net baryon number, independent control of the staggered potential and interaction--gradient detuning gives access to the three regimes illustrated in Fig.~\ref{fig:encoding}b.
Mesons are resonant with the vacuum at $m=3g^2/16$ and with baryon--antibaryon pairs at $m=-3g^2/16$.
Between these resonances, the mesonic configuration is energetically favored in the zero-tunneling limit.
The ideal mesonic state consists of a close-packed array of quark--antiquark pairs, with distinct gauge-link configurations within and between the pairs.
Outside this interval, negative mass favors a baryonic state with baryons and antibaryons alternating on every matter site, whereas positive mass favors the bare vacuum, with no matter excitations in this zero-tunneling limit.
The staggered polarization operator $\hat\Sigma=L^{-1}\sum_{\ell,c}(-1)^\ell\hat\psi_{\ell c}^{\dagger}\hat\psi_{\ell c}$ measures the normalized sublattice imbalance of the fermion occupation.
It takes the ideal values $+1$, $0$ and $-1$ for baryonic, mesonic and vacuum configurations, respectively.
Here $L$ is the number of matter sites, with even sites chosen as the particle sublattice.

We distinguish these hadronic configurations using a Stern--Gerlach-type spin readout combined with site-resolved fluorescence imaging (Fig.~\ref{fig:encoding}c) \cite{zhang2023scalable}.
After freezing the dynamics, we use a magnetic-field gradient in a transverse superlattice to separate the $\alpha$ and $\beta$ components into neighboring readout rows before pinning the atoms for imaging.
This spin-to-position mapping resolves the occupation of each spin component, distinguishing empty sites, the two singly occupied $\alpha$ or $\beta$ states, and $\alpha \beta$ doublons.
In the staggered-fermion encoding, doublons and holes on the particle and antiparticle sublattices represent baryons ($B$) and antibaryons ($\bar B$), respectively, whereas the same occupations on the opposite sublattices encode vacuum sites ($\Omega_d$ and $\Omega_h$).
In the atomic encoding, mesons correspond to nearest-neighbor atomic pairs (dimers) with a fixed spin ordering.
Mesons on the two bond sublattices are labeled $M$ and $M'$, corresponding to pairing configurations related by a one-site translation.
Within the encoded subspace, this reconstruction converts each atomic snapshot into a site-resolved map of hadrons and gauge-link configurations (Fig.~\ref{fig:encoding}d).

\section{\label{sec:ramp}Probing the hadronic phase structure}
\begin{figure*}[t]
    \centering
    \includegraphics[width=\textwidth]{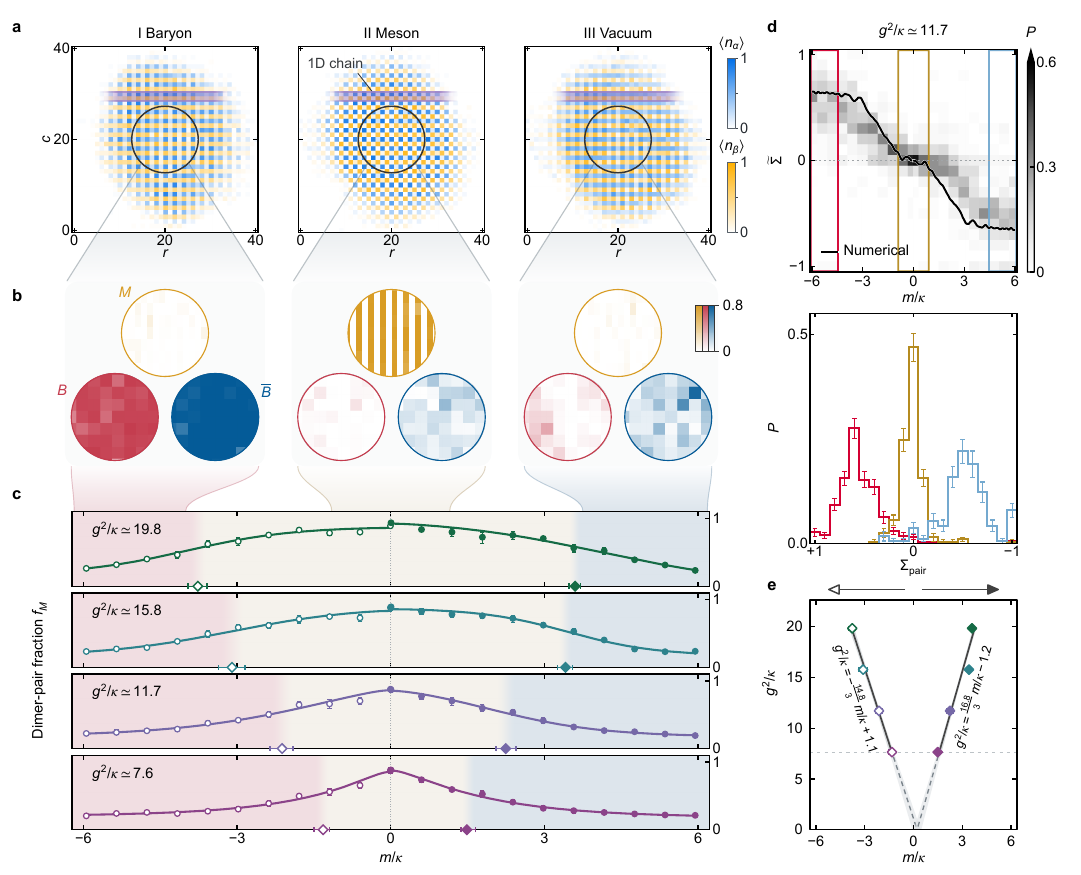}
    \caption{\textbf{Microscopic mapping of mesonic, baryonic and vacuum regimes.}
    \textbf{a,} Mean occupation maps after Stern--Gerlach readout in the baryonic (I), mesonic (II) and vacuum (III) regimes.
    Blue and gold denote the $\alpha$ and $\beta$ components, respectively; purple shading marks one chain's two readout rows.
    \textbf{b,} Meson ($M$), baryon ($B$) and antibaryon ($\bar B$) densities reconstructed from the circled regions in \textbf{a}.
    \textbf{c,} Meson-pair fraction $f_M$ during 100-ms mass ramps at four fixed gauge couplings.
    Open and filled circles denote negative- and positive-mass ramps, respectively.
    Curves are empirical sigmoidal fits; diamonds mark experimental transition points defined by the midpoint between fitted endpoint fractions.
    \textbf{d,} Distributions of normalized staggered polarization $\widetilde{\Sigma}$ at $g^2/\kappa=11.7(8)$ (top) and histograms for the marked mass windows (bottom).
    Black curve: matrix-product-state simulations of the two-component Bose--Hubbard model with $L=20$.
    \textbf{e,} Experimental transition points from \textbf{c} in the mass--coupling plane.
    Colors match \textbf{c}; solid lines are linear fits and dashed lines their extrapolations.
    Error bars denote 95\% confidence intervals in \textbf{c,e} and 68.27\% in \textbf{d}.
    Parentheses indicate uncertainties in the last quoted digits.
    }
    \label{fig:phase_mapping}
\end{figure*}

\begin{figure*}[t]
    \centering
    \includegraphics[width=\textwidth]{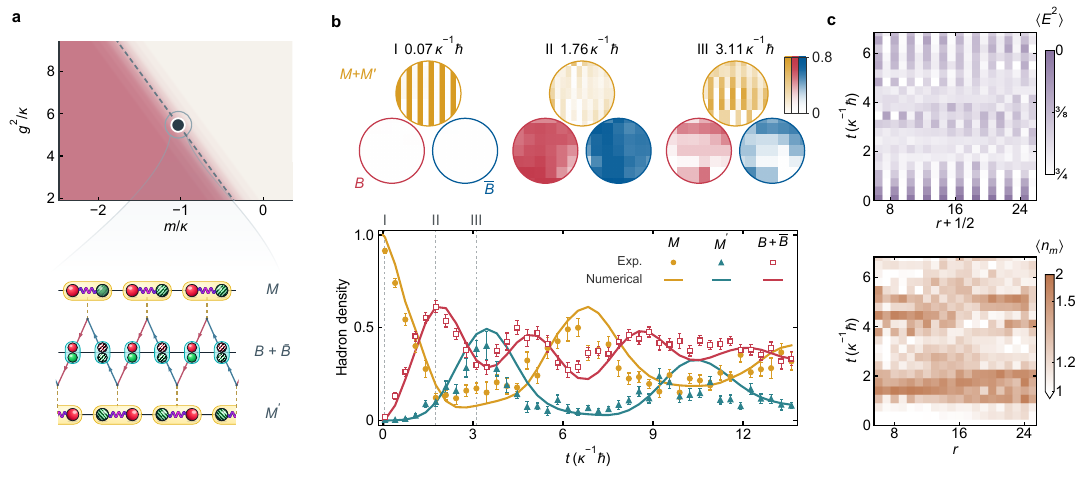}
    \caption{\textbf{Microscopic observation of meson--baryon oscillations.}
    \textbf{a,} Working point in the mass--gauge-coupling plane, close to the bare meson--baryon resonance $g^2=-\frac{16}{3}m$ (dashed line).
    The lower schematic shows meson-to-baryon--antibaryon conversion and meson regeneration on shifted bonds.
    \textbf{b,} Dynamics at $R$.
    Lower: mean site fractions of initial-bond mesons $M$ (gold circles), shifted-bond mesons $M'$ (teal triangles), and baryons plus antibaryons $B+\bar B$ (red squares).
    The two meson-bond sublattices differ by one lattice spacing.
    Error bars denote bootstrap standard errors.
    Solid lines show matrix-product-state simulations of the full two-component Bose--Hubbard model (BHM) on an open $L=20$ chain.
    The remaining site fraction includes encoded vacuum.
    Upper: local meson, baryon and antibaryon density maps in the same illustrative region at the marked times I--III.
    The meson maps show the combined population $M+M'$ on both bond sublattices.
    \textbf{c,} Space--time maps of the link Casimir $\langle E^2\rangle$ (upper) and matter excitation density $\langle n_m\rangle$ (lower), reconstructed from the same spin-resolved readout.
    Data in \textbf{b,c} correspond to zero net baryon number.
    }
    \label{fig:mb_dynamics}
\end{figure*}

We explore the three hadronic regimes by slowly ramping the mass from an initial mesonic configuration.
The experiment is performed in parallel on an array of one-dimensional chains, each typically about 20 sites long, with about 230 atoms in total.
Neighboring chains are separated by an initially empty lattice row, which provides auxiliary sites for Stern--Gerlach readout.
Starting from an alternating-spin meson configuration near $m=0$, we sweep the staggered potential linearly toward negative or positive mass over $100\,\mathrm{ms}$, corresponding to $20.6(4)\hbar/\kappa$, while keeping $\kappa/h=32.8(6)\,\mathrm{Hz}$ and $g^2$ fixed.
During Stern--Gerlach readout, one spin component is displaced into the intervening rows.
This maps the alternating occupations along the meson chains onto the two-dimensional checkerboard pattern in Fig.~\ref{fig:phase_mapping}a.
By contrast, doublons predominantly occupy the particle and antiparticle sublattices in the baryonic and vacuum regimes, respectively, producing striped readout patterns offset by one lattice site along the chains.
The microscopic readout separately resolves the local densities of the three hadronic components (Fig.~\ref{fig:phase_mapping}b).
These maps reveal meson-dominated populations in the mesonic regime and abundant baryons and antibaryons in the baryonic regime, while all three hadron densities are comparatively low in the vacuum regime.

To quantify this conversion, we measure the meson-pair fraction $f_M=D/(D+Q_{\rm p}+Q_{\rm a})$ during the slow mass ramps at different gauge couplings (Fig.~\ref{fig:phase_mapping}c).
Here $D$ is the total number of identified dimers on both bond sublattices, and $Q_{\rm p,a}$ are the doublon counts on the particle and antiparticle sublattices, respectively.
An ideal mesonic configuration gives $f_M=1$, whereas the ideal baryonic and vacuum configurations give $f_M=0$.
For each ramp, we identify the experimental transition point where the fitted meson fraction reaches the midpoint between its initial and final values.
Increasing $g^2/\kappa$ shifts these points to larger mass magnitudes and broadens the meson-dominated interval.
We map these transition points as a function of $g^2/\kappa$ (Fig.~\ref{fig:phase_mapping}e).
The measured boundaries follow the overall trends of the meson--baryon and meson--vacuum resonances at $g^2=-\frac{16}{3}m$ and $g^2=\frac{16}{3}m$, respectively.

The meson fraction alone does not distinguish the baryonic and vacuum regimes.
We therefore measure the normalized staggered polarization \mbox{$\widetilde{\Sigma}=(Q_{\rm p}-Q_{\rm a})/(D+Q_{\rm p}+Q_{\rm a})$}.
At $g^2/\kappa=11.7(8)$, the mass-resolved distribution is centered near zero in the mesonic regime, shifting to positive values in the baryonic regime and negative values in the vacuum regime (Fig.~\ref{fig:phase_mapping}d, upper).
The corresponding histograms from the three highlighted mass windows show a narrow mesonic peak around zero, whereas the baryonic and vacuum distributions are broader and centered on opposite sides of zero (Fig.~\ref{fig:phase_mapping}d, lower).
The meson fraction and staggered polarization jointly distinguish the three hadronic regimes.

\section{\label{sec:quench}Meson--baryon oscillations}

\begin{figure*}[t]
    \centering
    \includegraphics[width=\textwidth]{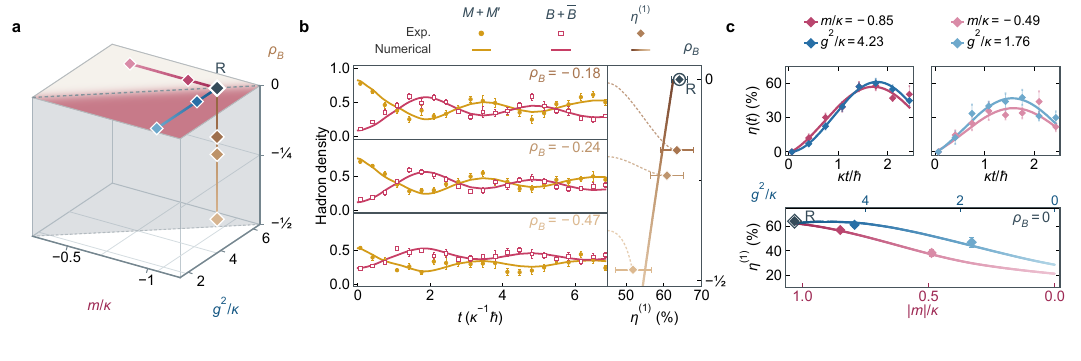}
    \caption{\textbf{Density and energy dependence of meson--baryon conversion.}
    \textbf{a,} Schematic density (copper), mass (red) and gauge-coupling (blue) directions from the shared reference $R$ of Fig.~\ref{fig:mb_dynamics}.
    The gray plane indicates the bare meson--baryon resonance $g^2=-16m/3$.
    \textbf{b,} Density dependence at the reference Hamiltonian.
    Left, total mesonic site fractions $M+M'$ (gold circles) and baryon-plus-antibaryon site fractions $B+\bar B$ (red squares) for three measured-density groups, with mean densities $\rho_B$ indicated.
    Right, first-peak conversion efficiencies $\eta^{(1)}$ versus net baryon density (copper diamonds), including the zero-density reference $R$.
    Dashed arrows connect the population traces to their corresponding first-peak efficiencies.
    \textbf{c,} Mass and gauge-coupling dependence at $\rho_B=0$.
    Upper, conversion dynamics $\eta(t)$ at two masses (red diamonds) and two gauge couplings (blue diamonds), with solid lines showing experimental fits.
    Darker and lighter shades indicate settings closer to and farther from $R$, respectively.
    Lower, measured first-peak efficiencies along the mass (bottom axis, red) and gauge-coupling (top axis, blue) scans, sharing the reference $R$.
    The mass scan fixes $g^2/\kappa=5.47(17)$ and the coupling scan fixes $m/\kappa=-1.03(2)$.
    Symbols show experimental data. Error bars denote bootstrap standard errors.
    All numerical curves are obtained from matrix-product-state simulations of the full two-component Bose--Hubbard model on open $L=20$ chains.
    }
    \label{fig:hadron_response}
\end{figure*}

The competition between quark-mass and color-electric energies that shapes the hadronic regimes also governs the dynamical conversion between mesonic and baryonic configurations \cite{calajo2024digital}.
To probe this conversion, we prepare a mesonic initial state and rapidly switch the lattice parameters to the working point $R$ (Fig.~\ref{fig:mb_dynamics}a).
The parameters are $m/\kappa=-1.03(2)$, $g^2/\kappa=5.47(17)$ and $\kappa/h=107.8(14)\,\mathrm{Hz}$.
Near resonance, the changes in mass and color-electric energy balance each other, allowing conversion into baryon--antibaryon pairs while the vacuum channel remains detuned \cite{kasper2020jaynes}.
This conversion creates a quark--antiquark pair without changing net baryon number; the reverse process can regenerate mesons on either bond sublattice.

At zero net baryon number, we resolve mesons on the initial bonds ($M$) and on bonds shifted by one lattice site ($M'$) (Fig.~\ref{fig:mb_dynamics}b).
Depletion of $M$ accompanies a rise in the baryon-plus-antibaryon site fraction to $0.61(3)$ at $t=1.76(2)\,\hbar/\kappa$ (II).
As the baryonic population subsequently falls, mesons re-form mainly on the shifted bonds.
At $t=3.11(4)\,\hbar/\kappa$ (III), $M'$ reaches a site fraction of $0.39(4)$, compared with $0.17(3)$ for $M$.
Regenerated mesons therefore occupy bonds connecting neighboring initial meson pairs.
Simulations of the full Bose--Hubbard model on $L=20$ sites also capture this delayed growth of $M'$.
Local density maps at times I--III visualize the corresponding changes in hadronic composition and the shift of regenerated meson bonds.

We reconstruct the matter excitation density $\langle n_m\rangle$ and the link Casimir $\langle E^2\rangle$ from the same spin-resolved snapshots (Fig.~\ref{fig:mb_dynamics}c).
Within the encoded sector, a local meson-to-baryon--antibaryon conversion increases the matter excitation number from two to four and changes the intervening link from $E^2=0$ to $3/4$.
Pair annihilation reverses these changes and restores low-Casimir links on the bonds where mesons form.
The space--time maps track this exchange between matter excitations and color-electric flux through successive conversion cycles.

\section{\label{sec:response}Density and energy dependence}

A conserved net baryon number does not fix the hadronic composition: mesons can convert into baryon--antibaryon pairs even in the presence of a net baryon imbalance.
We now ask how finite net baryon density affects this conversion, alongside the energy scales set by mass and gauge coupling \cite{kasper2020jaynes,cataldi2026string}.
We define $\rho_B=2(N_B-N_{\bar B})/L$ as the net baryon number per two matter sites, where $N_B$ and $N_{\bar B}$ are the baryon and antibaryon numbers in a chain of $L$ matter sites.
Figure~\ref{fig:hadron_response}a shows the three directions explored from $R$: ensembles with different measured net baryon densities at the reference Hamiltonian, and mass or gauge-coupling scans at zero net baryon density.

At the reference mass and gauge coupling, meson--baryon oscillations persist at finite net baryon density (Fig.~\ref{fig:hadron_response}b).
To compare pair production across densities, we use the conversion efficiency $\eta(t)=[p(t)-p(t_0)]/\rho_M(t_0)$.
Here $p(t)$ is the baryonic site fraction excluding excess baryons or antibaryons, and $\rho_M(t_0)$ is the initial mesonic site fraction.
The first maximum, $\eta^{(1)}$, falls from $(64.1\pm2.0)\%$ at $\rho_B=0$ to $(51.8\pm4.8)\%$ at $\rho_B\simeq-0.47$.
Both experiment and the Bose--Hubbard simulations suggest that a larger net baryon imbalance reduces the conversion response per initial meson, an effect associated with meson regeneration involving background antibaryons \hyperref[sm:1d:density_energy]{(Supplementary Information)}.

At zero net baryon density, we then vary the energies controlling conversion (Fig.~\ref{fig:hadron_response}c).
Mesonic configurations couple to baryon--antibaryon configurations through quark--antiquark pair creation, and to the encoded vacuum through pair annihilation.
The corresponding bare energy differences are $3g^2/8+2m$ and $3g^2/8-2m$, respectively.
Changing the mass shifts these energies in opposite directions, whereas changing the gauge coupling shifts their common color-electric contribution.
Across both scans, the measured first-peak efficiencies decrease as $|m|/\kappa$ or $g^2/\kappa$ is reduced from $R$.
The Bose--Hubbard simulations capture the overall energy dependence of the measured conversion response.
Net baryon density and competing mass and color-electric energies shape the observed conversion response \hyperref[sm:1d:density_energy]{(Supplementary Information)}.

\section{\label{sec:discuss}Discussion and Outlook}

We have realized an analog simulator of a truncated one-dimensional SU(2) gauge theory with microscopic access to hadrons and color-electric flux.
Tunable mass and gauge coupling connect distinct hadronic regimes to real-time meson--baryon conversion and meson regeneration on shifted bonds.
The simulator provides access to mesons, color-singlet baryons and their interconversion at finite net baryon density.
Site- and bond-resolved measurements connect local meson--baryon conversion to collective hadron dynamics and the accompanying evolution of color-electric flux.

This platform can test whether reversible hadron conversion leads to chemical equilibration, a question relevant to heavy-ion collisions \cite{pan2014baryon}.
Preparing different hadronic configurations at fixed energy and net baryon number would reveal whether late-time populations lose memory of their initial state \cite{zhou2022thermalization,calajo2025scarring,ciavarella2025fragmentation}.
New atomic encodings and gauge-invariant couplings could extend this approach to higher dimensions and the SU(3) color symmetry of QCD \cite{banerjee2013atomic,wiese2022quantum,halimeh2024spin,depaciani2026fermionic}.
Such extensions would enable microscopic studies of hadron formation and nonequilibrium strong-interaction dynamics.

\begin{acknowledgments}
We thank Ying-Ying Li for carefully proofreading the manuscript and Peter Zoller for helpful discussions.
This work was supported by NSFC Grant No. 12125409 and the Innovation Program for Quantum Science and Technology 2021ZD0302000. 
W.-Y.Z.~acknowledges support from NSFC under Grant No. 12404323 and No. 12674336, the Fundamental Research Funds for the Central Universities under Grant No. WK2030000104. 
Z.-Y.Z.~acknowledges support from NNSFC under Grant No. 12504310, the 77th Batch of the General Program of the China Postdoctoral Science Foundation (2025M773437), and the Postdoctoral Fellowship Program of CPSF (GZC20252263).

\end{acknowledgments}

\par\endgroup
\bibliography{su2}
\bibliographystyle{naturemag}

\hypersetup{allcolors=black}
\onecolumngrid
\vspace*{0.5cm}
\newpage
\begin{center}
    \textbf{METHODS AND SUPPLEMENTARY MATERIALS}
\end{center}
\vspace*{0cm}

\incltocpage
\tableofcontents
\appendix
\counterwithout{equation}{section}

\setcounter{equation}{0}
\setcounter{figure}{0}

\setcounter{secnumdepth}{3}
\makeatletter
\makeatother

\newcommand{\lk}[1]{\ket{\mathbf{#1}}}

\renewcommand{\theequation}{S\arabic{equation}}
\renewcommand{\theHequation}{supp.\arabic{equation}}
\renewcommand{\figurename}{Fig.}
\renewcommand{\thefigure}{S\arabic{figure}}
\renewcommand{\theHfigure}{supp.\arabic{figure}}
\renewcommand{\thetable}{S\arabic{table}}
\renewcommand{\thesection}{S\arabic{section}}
\renewcommand{\thesubsection}{\Alph{subsection}}
\renewcommand{\thesubsubsection}{\arabic{subsubsection}}

\newpage

\section{Experimental implementation}
\label{sm:experiment}

\subsection{Experimental system}
\label{sm:setup}

\subsubsection{Setup}
\label{sm:setup:apparatus}

We prepare a two-dimensional (2D) Bose-Einstein condensate of $^{87}$Rb atoms in the $5S_{1/2},~\ket{F=1,m_F=-1}$ state, confined to a single antinode of a vertical $1064$-nm lattice~\cite{zhang2023scalable}.
The vertical-lattice beam has a $1/e^2$ intensity radius of $103\,\mu\mathrm{m}$.
Its Gaussian envelope provides smooth confinement in the horizontal plane, including along the chains, and produces the same potential for the two internal states used in the simulator.
The trap frequency near the cloud center is $46.0(4)\,\mathrm{Hz}$, determined from breathing-mode oscillations as described in Sec.~\ref{sm:calib:envelope}.
In the horizontal plane, pairs of $532$-nm laser beams intersect at approximately $50^{\circ}$ to form blue-detuned short lattices along $x$ and $y$, with spacing $a_s=630$~nm.
Red-detuned long lattices with spacing $a_l=2a_s$ are formed by $1064$-nm beams along the same optical paths~\cite{zhang2023scalable}.
The superposition of short and long lattices produces the optical superlattice potentials
\begin{align}
V(x)=V_{s,x}\cos^2(k_sx)-V_{l,x}\cos^2(k_lx+\theta_x), \nonumber \\
V(y)=V_{s,y}\cos^2(k_sy)-V_{l,y}\cos^2(k_ly+\theta_y),
\label{eq:superlattice}
\end{align}
where $V_{s,x(y)}$ and $V_{l,x(y)}$ are the depths of the short and long lattices, $k_s=\pi/a_s$ and $k_l=\pi/a_l$ are the corresponding wavevectors, and $\theta_{x(y)}$ is their relative phase, controlled using galvanometers~\cite{zhang2023scalable}.

The polarization of the long-lattice beams controls the spin dependence of the superlattice and enables parallel, sublattice-selective spin manipulation~\cite{yang2017spin,zhang2023scalable}.
Our previous setup used electro-optic modulators (EOMs) to control the polarization~\cite{zhang2023scalable}.
Here, half-wave plates mounted on motorized galvanometers replace the EOMs in both long-lattice beam paths.
Rotation of the wave plates sets the polarization of the transmitted beams.

The higher optical power supported by this arrangement allows larger beam waists at the required lattice depths.
The long-lattice beams have elliptical waists of approximately $130~\mu$m in the horizontal plane and $65~\mu$m vertically, compared with about $50~\mu$m in the previous setup.
The larger waists improve the uniformity of the lattice potential across the atomic cloud, while wave-plate control reduces spatial variations of the polarization.

Atomic fluorescence is collected through a high-numerical-aperture objective ($\mathrm{NA}=0.8$) to enable site-resolved detection of atomic occupations.

The one-dimensional (1D) chains lie along $y$, and the long lattice along this direction produces the staggered offset $\delta$ of Sec.~\ref{sm:setup:bhm}.
During the mass ramps and quenches, the long-lattice polarization is set to give the same staggered offset for both spin states.

\subsubsection{Spin-dependent gradient}
\label{sm:setup:gradient}

In the 1D experiments, we apply a magnetic-field gradient $\partial_y B$ along the chains on top of a homogeneous bias field $B_0 \approx 5~\mathrm{G}$.
The two internal states used in the experiments, $\alpha \equiv \ket{F=1,m_F=-1}$ and $\beta \equiv \ket{F=1,m_F=+1}$, belong to the $F=1$ manifold with Land\'e factor $g_F = -1/2$.
Their linear Zeeman shifts $E_{m_F} = g_F m_F \mu_B B$ are equal in magnitude and opposite in sign (see Sec.~\ref{sm:states:zeeman}).
The gradient therefore produces opposite tilts for the two states, with a nearest-neighbor energy offset of magnitude
\begin{equation}
    \Delta = \lvert g_F \rvert  \mu_B \lvert \partial_y B \rvert a_s,
    \label{eq:gradient}
\end{equation}
where $\lvert m_F \rvert = 1$ for both states and $a_s$ is the short lattice spacing.
For the chosen orientation of the magnetic gradient, the energy of state $\alpha$ decreases by $\Delta$ per site along the $+y$ direction, whereas that of state $\beta$ increases by the same amount; these are the opposite offsets $\mp \Delta$ of the main text.
For $a_s=630$~nm, a gradient of $1\,\mathrm{G\,cm^{-1}}$ gives $\Delta/h\simeq44$~Hz, where $h$ is the Planck constant.
The gradient $\Delta$ is calibrated independently as described in Sec.~\ref{sm:calib:UV}.

\subsubsection{Two-component Bose-Hubbard model}
\label{sm:setup:bhm}

For a 1D chain of $L$ sites with open boundary conditions, we label the sites by $\ell = 0, 1, \ldots, L-1$, with $\ell = 0$ an even site.
As in the main text, the even sites form the particle sublattice and the odd sites form the antiparticle sublattice of the staggered fermion encoding (Sec.~\ref{sm:lgt}).
Then, the atomic system is described by the two-component Bose-Hubbard Hamiltonian
\begin{equation}
    \begin{split}
        \hat H_{\mathrm{BH}} = {} & -J \sum_{\ell=0}^{L-2} \sum_{s=\alpha,\beta}
        \left( \hat s^\dagger_\ell \hat s_{\ell+1} + \mathrm{h.c.} \right)
        + \frac{U}{2} \sum_{\ell,s} \hat n_{\ell,s} \left( \hat n_{\ell,s} - 1 \right) \\
        {} & + V \sum_\ell \hat n_{\ell,\alpha} \hat n_{\ell,\beta}
        + \Delta \sum_\ell \ell \left( \hat n_{\ell,\beta} - \hat n_{\ell,\alpha} \right)
        + \delta \sum_{\ell,\mathrm{even}} \hat n_\ell ,
    \end{split}
    \label{eq:bhm}
\end{equation}
where $\hat n_{\ell,s}=\hat s^\dagger_\ell\hat s_\ell$ and $\hat n_\ell=\hat n_{\ell,\alpha}+\hat n_{\ell,\beta}$.
Here, $J$ is the nearest-neighbor tunneling amplitude, taken to be identical for the two components; $U$ and $V$ are the intra- and inter-component on-site interaction energies; $\Delta$ is the spin-dependent nearest-neighbor energy offset introduced in Eq.~\eqref{eq:gradient}; and $\delta$ is the staggered superlattice offset produced by the long lattice.
With the convention used here and in the main text, $\delta>0$ ($\delta<0$) raises (lowers) the energy of the even (particle) sites relative to the odd (antiparticle) sites; and the staggered offset sets the quark mass, $m = \delta / 2$ (Sec.~\ref{sm:lgt}).
For a fixed total atom number, the staggered term can equivalently be written, up to an additive constant, as
\begin{equation}
    \delta \sum_{\ell,\mathrm{even}} \hat n_\ell \equiv \frac{\delta}{2} \sum_\ell (-1)^\ell \hat n_\ell + \mathrm{const.} = \frac{\delta L}{2} \hat \Sigma + \mathrm{const.},
    \label{eq:staggered}
\end{equation}
where $\hat \Sigma = L^{-1} \sum_\ell(-1)^\ell\hat n_\ell$ is the staggered atomic polarization operator of the main text.
All energy scales are quoted as frequencies $E/h$ in units of Hz unless stated otherwise.
The numbers $N_\alpha$ and $N_\beta$ of the atoms in the two states are conserved separately (Sec.~\ref{sm:states}), and all experiments are performed at $N_\alpha = N_\beta = N/2$.

\subsubsection{Experimental sequence}
\label{sm:setup:sequence}

The experimental sequence consists of initial-state preparation, Hamiltonian evolution and spin-resolved detection.

\begin{figure}[H]
    \centering
    \includegraphics[width=\textwidth]{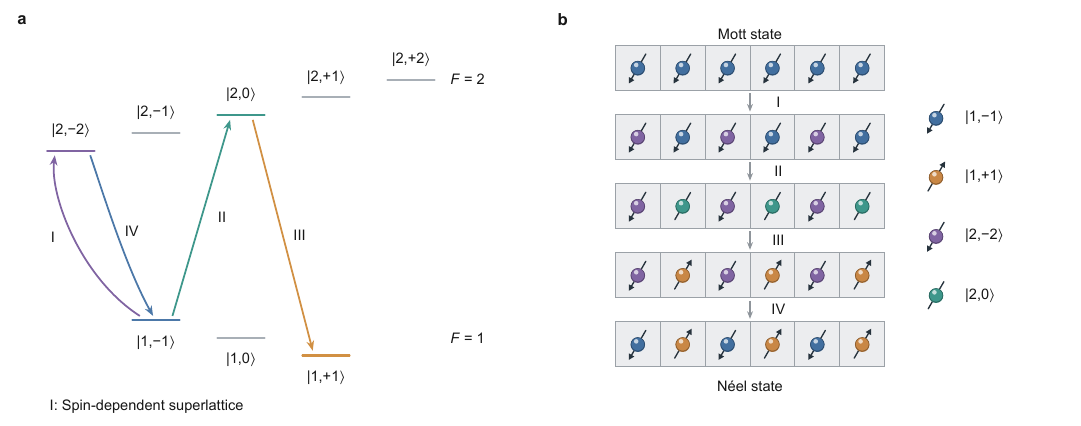}
    \caption{\textbf{Preparation of the atomic N\'eel state.}
    \textbf{a,} Ground-state hyperfine Zeeman levels and four microwave Landau--Zener transfers, I--IV.
    The spin-dependent superlattice makes step I selective to one sublattice.
    Energy-level spacings are schematic.
    \textbf{b,} Atomic configurations through the sequence for three neighboring unit cells.
    Gray cells mark lattice sites; atom colors identify the internal states labeled in \textbf{a}.
    The sequence converts a polarized Mott insulator into an atomic N\'eel state.}
    \label{fig:neel_preparation}
\end{figure}

\begin{description}
\item[(i)] \emph{Mott insulator preparation.}
We prepare Mott-insulating chains using staggered-immersion cooling~\cite{zhang2023scalable}.
Before the superfluid-to-Mott-insulator transition, a long lattice along $x$ divides the system into alternating chains.
After cooling, the reservoir chains are removed, leaving parallel chains of $\alpha$ atoms along $y$ separated by empty rows.
These rows provide the auxiliary sites for the subsequent spin-resolved readout.

\item[(ii)] \emph{N\'eel-state preparation.}
Starting from one $\alpha$ atom per occupied site, we prepare the alternating spin pattern through four microwave Landau--Zener transfers (Fig.~\ref{fig:neel_preparation}).
The spin-dependent superlattice along $y$ shifts the microwave transition frequencies differently on the two sublattices, providing spatial selectivity for the first transfer.
In step I, atoms on sublattice A are temporarily transferred from $\alpha=\ket{1,-1}$ to $\ket{2,-2}$.
The remaining $\alpha$ atoms on sublattice B are then transferred to $\ket{2,0}$ in step II and to $\beta=\ket{1,+1}$ in step III.
In step IV, the atoms stored in $\ket{2,-2}$ are returned to $\alpha$.
Each transfer uses a microwave frequency sweep through the corresponding transition.
The resulting atomic N\'eel state is $\ket{\alpha,\beta,\alpha,\beta,\dots}$, with A and B corresponding to the even and odd sites in our convention.
The same sequence prepares the parallel chains simultaneously.

\item[(iii)] \emph{Dynamics.}
For the mass-ramp measurements, we set $J$ and $\Delta$ to their chosen values and sweep the staggered offset linearly from zero to $\delta_{\mathrm{end}}/h=\pm390$~Hz over $100$~ms.
The magnetic-field gradient and the short-lattice intensity remain fixed throughout each ramp.
For the quench measurements, we first set $\Delta$ and $\delta$ to their chosen values in a deep lattice, where tunneling is negligible.
We then lower the lattice along the chains in $0.1\,\mathrm{ms}$, increasing $J$ to its final value and initiating the dynamics.
The evolution time $t$ is measured from the end of this lattice ramp.

\item[(iv)] \emph{Detection.}
After the evolution time $t$, the lattices are ramped up to freeze the dynamics, and the state is read out as described in Sec.~\ref{sm:setup:readout}.

\end{description}

\subsubsection{Site- and spin-resolved readout}
\label{sm:setup:readout}

Fluorescence imaging records the on-site occupation parity through light-assisted pair loss~\cite{bakr2009quantum,sherson2010single}.
We obtain spin-resolved information by first mapping the two internal states onto neighboring readout sites, using the empty rows left by staggered-immersion cooling.

After freezing the dynamics along $y$, we rotate the magnetic-field gradient to the transverse $x$ direction.
The optical superlattice along $x$ forms double wells connecting each occupied chain to its neighboring empty row.
The gradient and superlattice produce a Stern--Gerlach-type separation of the $\alpha$ and $\beta$ components into opposite wells~\cite{zhang2023scalable}.
Each original lattice site is thus mapped onto two spatially resolved readout sites whose positions distinguish the internal states.

For an on-site $\alpha\beta$ pair, the separation converts the readout-well occupations from $(2,0)$ to $(1,1)$, allowing both atoms to be imaged individually.
The resolved readout pattern gives the complete spin-resolved occupation configuration after separation.
Using the spin-to-position mapping, we reconstruct $(n_{\ell,\alpha},n_{\ell,\beta})$ on the original chain and assign the corresponding hadronic configurations using Table~\ref{tab:basis}.
As illustrated in Fig.~\ref{fig:encoding}d, each site or bond is identified as a meson ($M$ or $M'$), a baryon ($B$), an antibaryon ($\bar B$) or a vacuum site ($\Omega_d$ or $\Omega_h$).

\subsection{Choice of the two internal states}
\label{sm:states}

We choose the two internal states
\begin{equation}
    \alpha\equiv\ket{F=1,m_F=-1},\qquad \beta\equiv\ket{F=1,m_F=+1}
\end{equation}
to combine equal and opposite linear Zeeman shifts, nearly equal interaction strengths and long-lived on-site pairs.
These properties support spin-dependent resonant tunneling and the stable doublons required by the atomic encoding.

\subsubsection{Equal and opposite magnetic moments}
\label{sm:states:zeeman}

Both states belong to the $F=1$ hyperfine manifold.
To first order in the magnetic field, their Zeeman shifts are $E_{m_F}=g_Fm_F\mu_BB$ with $g_F\simeq-1/2$; since $m_F=\mp1$ for $\alpha$ and $\beta$, the products $g_Fm_F$ are equal in magnitude and opposite in sign.
A magnetic-field gradient therefore produces the opposite tilts $-\Delta$ and $+\Delta$ per site in Eq.~\eqref{eq:bhm}.
The quadratic Zeeman shift $q\,m_F^2$ is common to the two states.
For comparison, the combination $\ket{F=1,m_F=-1}$ and $\ket{F=2,m_F=-1}$ also has approximately equal and opposite first-order magnetic moments, whereas the magnetic moment of $\ket{F=2,m_F=-2}$ is about twice as large in magnitude as that of $\alpha$.
The lifetime comparison in Sec.~\ref{sm:states:lifetime} provides an additional basis for choosing the two $F=1$ states.

\subsubsection{Collisional properties}
\label{sm:states:collisions}

For two $^{87}$Rb atoms in the $F=1$ manifold, $s$-wave collisions proceed in the symmetric total-spin channels $F_{\mathrm{tot}}=0$ and $2$, with scattering lengths $a_0=101.8\,a_B$ and $a_2=100.4\,a_B$, where $a_B$ is the Bohr radius~\cite{vanKempen2002}.
The same-component pairs $\alpha\alpha$ and $\beta\beta$ belong to the $F_{\mathrm{tot}}=2$ channel, so that $a_{\alpha\alpha}=a_{\beta\beta}=a_2$ exactly.
The symmetrized interspecies state decomposes as
\begin{equation}
    \ket{\alpha\beta}_{\mathrm{sym}}=\sqrt{\tfrac13}\,\ket{F_{\mathrm{tot}}=2,M_F=0}+\sqrt{\tfrac23}\,\ket{F_{\mathrm{tot}}=0,M_F=0},
\end{equation}
with $M_F$ the total spin projection, and its elastic scattering length is
\begin{equation}
    a_{\alpha\beta}=\frac{a_2+2a_0}{3}\simeq101.3\,a_B .
    \label{eq:interspecies-scattering}
\end{equation}
For normalized Wannier functions $w_\alpha$ and $w_\beta$, the single-band interaction energies are
\begin{equation}
    U_s=\frac{4\pi\hbar^2a_{ss}}{M_{\mathrm{Rb}}}\int d^3r\,\lvert w_s(\mathbf r)\rvert^4\quad(s=\alpha,\beta),\qquad
    V=\frac{4\pi\hbar^2a_{\alpha\beta}}{M_{\mathrm{Rb}}}\int d^3r\,\lvert w_\alpha(\mathbf r)\rvert^2\lvert w_\beta(\mathbf r)\rvert^2 ,
    \label{eq:interaction-overlaps}
\end{equation}
where $M_{\mathrm{Rb}}$ is the atomic mass.
For identical Wannier functions, $U_\alpha=U_\beta\equiv U$ and
\begin{equation}
    \frac VU=\frac{a_{\alpha\beta}}{a_2}\simeq1.009 .
    \label{eq:interaction-ratio}
\end{equation}
The scattering lengths therefore give nearly equal intra- and inter-component interactions for common spatial wavefunctions, consistent with $V\simeq U$ in Sec.~\ref{sm:lgt:validity}.
The overlap integrals in Eq.~\eqref{eq:interaction-overlaps} specify the interaction energies when the spatial wavefunctions differ.
Keeping both atoms in the lower hyperfine manifold also removes the energy release associated with relaxation from $F=2$ to $F=1$.

\subsubsection{Suppression of spin-changing collisions}
\label{sm:states:spinexchange}

Within the $F=1$ manifold, contact interactions couple an $\alpha\beta$ pair to two atoms in $\ket{F=1,m_F=0}$ through
\begin{equation}
    \ket{1,-1}\ket{1,+1}\longleftrightarrow\ket{1,0}\ket{1,0}.
    \label{eq:spinexchange}
\end{equation}
The coupling is proportional to the small scattering-length difference $a_2-a_0$~\cite{Widera2005PRL,Widera2006NJP}.
For a common spatial wavefunction, its matrix element is $\lambda_{\mathrm{ex}}=\sqrt{2}(a_2-a_0)U/(3a_2)$, giving $\lvert\lambda_{\mathrm{ex}}\rvert/h$ of a few hertz for $U/h$ near $1\,\mathrm{kHz}$.
The bias field adds a quadratic Zeeman energy difference $2q$ between the two pair states, with $q/h\simeq72~\mathrm{Hz\,G^{-2}}\times B^2$ for $^{87}$Rb in $F=1$.
At $B_0\simeq5$~G, this contribution is about $3.6$~kHz, much larger than the spin-exchange matrix element.
This energy separation suppresses spin exchange and supports the two-component description used in Eq.~\eqref{eq:bhm}.

\subsubsection{Lifetimes of on-site pairs}
\label{sm:states:lifetime}

\begin{figure}[htbp]
    \centering
    \includegraphics[width=89mm]{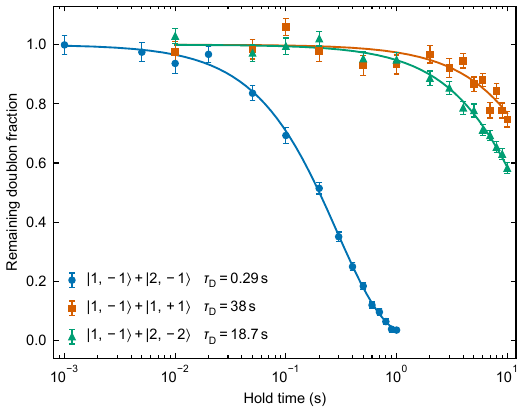}
    \caption{\textbf{Doublon lifetimes for three internal-state combinations.}
    Each doublon contains one atom in each indicated state, written as $\ket{F,m_F}$.
    Symbols show the normalized mean pair counts after spin separation; error bars denote one standard error, scaled by the fitted amplitude $A$.
    Solid lines show exponential fits, with fitted lifetimes $\tau_D$ given in the legend.
    The plotted fraction is $[\overline{N}_{11}(t)-C]/A$; the constant background is fitted for the blue series and fixed to zero for the orange and green series.
    The orange series is the $\alpha\beta$ pair used in the gauge simulator.}
    \label{fig:lifetime}
\end{figure}

We compare on-site pairs containing one atom in $\ket{F=1,m_F=-1}$ and a second atom in $\ket{F=2,m_F=-1}$, $\ket{F=1,m_F=+1}$ or $\ket{F=2,m_F=-2}$.
After a variable hold time in the optical lattice, spin separation maps each surviving pair onto a double well with occupation $(1,1)$, which is counted in the fluorescence image.
Each curve in Fig.~\ref{fig:lifetime} contains 15 hold times, with 62--67 independent realizations per time point.

We fit the mean pair count to $\overline{N}_{11}(t)=A\exp(-t/\tau_D)+C$ by weighted least squares, with weights equal to the inverse squared standard errors.
The fit includes a constant background $C$ for the $\ket{1,-1}$ and $\ket{2,-1}$ combination; for the other two combinations, $C=0$.
The plotted fraction is $[\overline{N}_{11}(t)-C]/A$.
Lifetime intervals are obtained from 3,000 bootstrap resamples of the experimental realizations, sampled separately at each hold time with the chosen fit model held fixed.

For the selected $\alpha\beta$ pair, the fitted lifetime is $\tau_D=37.9\,\mathrm{s}$, with a 95\% bootstrap interval of $32.0$--$45.9\,\mathrm{s}$.
The $\ket{1,-1}$ and $\ket{2,-1}$ combination has a decay time of $0.290\,\mathrm{s}$ ($0.269$--$0.314\,\mathrm{s}$), while the $\ket{1,-1}$ and $\ket{2,-2}$ combination gives $18.7\,\mathrm{s}$ ($17.4$--$20.3\,\mathrm{s}$), with the same interval convention.
The fitted $\alpha\beta$ lifetime is about 380 times the $100$-ms mass-ramp duration.
Together with its symmetric magnetic response, this long lifetime makes the two $F=1$ states well suited to the doublon-based encoding.

\subsection{Calibration of the Hubbard parameters}
\label{sm:calib}

We calibrate the lattice depths and the Hubbard parameters $J$, $U$, and $\Delta$ using the procedures developed in our previous work~\cite{zhang2023scalable,wang2023interrelated}.
The Bose--Hubbard parameters and vertical-lattice confinement for the mass ramps and reference quench are summarized in Table~\ref{tab:parameters}.

\begin{table}[htbp]
\caption{\textbf{Bose--Hubbard parameters and vertical-lattice confinement.}
Parameters for the mass ramps and quenches; the quench column lists the reference setting $R$ first, followed by the mass- and coupling-scan settings.
Parentheses give uncertainties in the last quoted digits.}
\label{tab:parameters}
\renewcommand{\arraystretch}{1.2}
\begin{ruledtabular}
\begin{tabular}{lccc}
Parameter & Unit & Mass ramp & Quench \\ \hline
Tunneling $J/h$ & Hz & $23.2(4)$ & $76.2(10)$ \\ \hline
Interaction $V/h\sim U/h$ & Hz & $894(10)$ & $821(6)$ \\ \hline
Spin-dependent tilt $\Delta/h$ & Hz & \begin{tabular}[c]{@{}c@{}}$650.0(13),\ 700.0(13),$\\$750.0(14),\ 800.0(15)$\end{tabular} & \begin{tabular}[c]{@{}c@{}}$600.0(12),$\\$650.0(13),\ 750.0(14)$\end{tabular} \\ \hline
Staggered offset $\delta/h$ & Hz & $0\to\pm390(5)$ & \begin{tabular}[c]{@{}c@{}}$-222.3(29),$\\$-183.3(24),\ -105.3(14)$\end{tabular} \\ \hline
Measured trap frequency $\nu_{\mathrm{eff}}$ & Hz & \multicolumn{2}{c}{$46.0(4)$} \\
\end{tabular}
\end{ruledtabular}
\end{table}

\subsubsection{Lattice depths and superlattice phases}
\label{sm:calib:lattice}

We calibrate the depths of the short, long, and vertical lattices by lattice-modulation spectroscopy: we modulate the lattice intensity sinusoidally, detect interband excitations by band mapping, and compare the resonance of the $s$--$d$ interband transition with the calculated band structure~\cite{zhang2023scalable}.
To calibrate the relative superlattice phase $\theta_{x(y)}$, atoms initially confined in the long-lattice wells are split into double wells by raising the short lattice, and the left--right population contrast $I=\lvert(n_L-n_R)/(n_L+n_R)\rvert$, with $n_L$ and $n_R$ the mean occupations of the two wells, is measured as a function of the galvanometer setting~\cite{zhang2023scalable}.
Balanced double wells define $\theta_{x(y)}=0$ by convention in Eq.~\eqref{eq:superlattice}.

\subsubsection{Confinement from the vertical-lattice envelope}
\label{sm:calib:envelope}

We calibrate the in-plane confinement produced by the Gaussian envelope of the vertical lattice using breathing-mode oscillations of the atomic cloud.
After evaporative cooling in a single layer, we abruptly release the atoms from the tightly focused dimple trap into the broader confinement of the vertical lattice and record the periodic expansion and contraction of the cloud as a function of hold time.
In the harmonic approximation, the breathing-mode frequency is twice the trap frequency.
This procedure gives a trap frequency of $46.0(4)\,\mathrm{Hz}$ for the vertical-lattice confinement.

\subsubsection{Tunneling amplitude $J$}
\label{sm:calib:J}

To calibrate the tunneling amplitude along $x$ or $y$, we isolate a single chain of atoms and initiate a quantum walk by rapidly lowering the lattice depth along the chain~\cite{zhang2023scalable,preiss2015strongly}.
We then record the site-resolved density distribution after a variable evolution time $\tau$.
For a nearest-neighbor chain with a uniform residual tilt $E_t$ per site, the probability of finding the atom
$\ell$ sites from its initial position is
\begin{equation}
p_\ell(\tau)=\mathcal J_\ell^{\,2}\!\left[\frac{4J}{E_t}\sin\!\left(\frac{E_t\tau}{2\hbar}\right)\right]
\ \xrightarrow{\ E_t\to0\ }\ \mathcal J_\ell^{\,2}\!\left(\frac{2J\tau}{\hbar}\right),
\label{eq:calib-quantum-walk}
\end{equation}
where $\mathcal J_\ell$ is the Bessel function of the first kind.
Fits to the spatial and temporal distributions determine $J$ and the residual tilt.
The calibrated amplitudes are $J/h=23.2(4)\,\mathrm{Hz}$ for the mass ramp and $J/h=76.2(10)\,\mathrm{Hz}$ for the quench.

\subsubsection{Interactions and spin-dependent gradient}
\label{sm:calib:UV}

We calibrate the intra-component interaction $U$ and the gradient $\Delta$ by modulation spectroscopy of a single-component Mott insulator in the tilted lattice: interaction-assisted tunneling produces resonances at the modulation frequencies $\Delta\pm U$, from which we extract both quantities~\cite {ma2011photon,zhang2023scalable}.
The inter-component interaction is taken as $V\simeq U$, in accordance with the scattering-length ratio of Eq.~\eqref{eq:interaction-ratio}.

\subsubsection{Staggered potential $\delta$}
\label{sm:calib:delta}

The staggered potential is controlled by the depth of the long lattice along the chains and is calibrated by resonant tunneling of isolated atoms in the superlattice, following Ref.~\cite{wang2023interrelated}.
We apply a magnetic energy bias $\Delta_B$ between the two sites of a double well, lower the barrier for a fixed duration, and measure the transferred population; tunneling is resonantly enhanced when the bias compensates the staggered offset, $\Delta_B=\delta$.
Repeating the measurement for different control voltages provides the conversion between voltage and $\delta$.
For the mass ramps, a control range of $\pm1\,\mathrm{V}$ gives endpoint offsets $\delta/h=\pm390(5)\,\mathrm{Hz}$.
For the quenches, long-lattice control voltages of $0.57$, $0.47$ and $0.27\,\mathrm{V}$ give $\delta/h=-222.3(29)$, $-183.3(24)$ and $-105.3(14)\,\mathrm{Hz}$, respectively.
The sign of $\delta$ is fixed by the sublattice on which doublons are energetically favored, consistent with the baryon- and vacuum-type configurations of Sec.~\ref{sm:lgt:crystals}.

\section{SU(2) lattice gauge theory and its atomic implementation}
\label{sm:lgt}

This section establishes the theoretical model and its connection to the atomic Hamiltonian.
We first define the representation-truncated SU(2) gauge theory, then describe the fixed Gauss-charge sector and its hadronic dynamics, and finally derive the atomic encoding and the conditions for its realization.
The algebraic correspondence between the two projected Hamiltonians is established here; simulations of the full atomic dynamics at the experimental parameters are addressed in Sec.~\ref{sm:1d}.

\subsection{SU(2) Hamiltonian and representation truncation}
\label{sm:lgt:qlm}

\subsubsection{Matter, gauge fields and local symmetry}

We consider an open chain of $L$ matter sites, numbered $\ell=0,\ldots,L-1$, with staggered fermions in the fundamental representation of SU(2).
The Hamiltonian is \cite{kogut1975hamiltonian,kuhn2015nonabelian}
\begin{equation}
\begin{split}
\hat H_{\mathrm{LGT}}={}&-\kappa\sum_{\ell=0}^{L-2}\sum_{c,c'}
\left(\hat\psi^\dagger_{\ell c}\hat U^{cc'}_\ell\hat\psi_{\ell+1,c'}+\mathrm{h.c.}\right)\\
&+m\sum_{\ell=0}^{L-1}(-1)^\ell\hat n^\psi_\ell
+\frac{g^2}{2}\sum_{\ell=0}^{L-2}\hat E_\ell^2,
\end{split}
\label{eq:lgt}
\end{equation}
where $c,c'\in\{\mathrm r,\mathrm g\}$, $\hat n^\psi_\ell=\sum_c\hat\psi^\dagger_{\ell c}\hat\psi_{\ell c}$, and link $\ell$ joins sites $\ell$ and $\ell+1$.
The local matter basis is $\{\ket0,\ket{\mathrm r},\ket{\mathrm g},\ket{\mathrm{rg}}\}$, with $\ket{\mathrm{rg}}=\hat\psi^\dagger_{\ell\mathrm r}\hat\psi^\dagger_{\ell\mathrm g}\ket0$.
Empty and doubly occupied matter states are color singlets, whereas the singly occupied states form a color doublet.
The link operator $\hat U_\ell$ transports color during matter hopping, and $\hat E_\ell^2$ is the quadratic electric-field Casimir.
In one spatial dimension the open chain has no magnetic plaquette term.

We use the same Gauss generators as in the main text,
\begin{equation}
\hat G^a_\ell=\hat Q^a_\ell+\hat R^a_{\ell-1}+\hat L^a_\ell,
\qquad
\hat Q^a_\ell=\frac12\sum_{c,c'}\hat\psi^\dagger_{\ell c}\sigma^a_{cc'}\hat\psi_{\ell c'}.
\label{eq:gauss_full}
\end{equation}
Here $\hat L^a$ and $\hat R^a$ act on the left and right ends of a link, and the exterior link ends specify the boundary electric fields.
The three generators obey $[\hat G^a_\ell,\hat G^b_k]=i\delta_{\ell k}\epsilon^{abc}\hat G^c_\ell$ and commute with $\hat H_{\mathrm{LGT}}$.
Thus $\hat{\mathbf G}_\ell^2$ and $\hat G^z_\ell$ label invariant sectors of the dynamics.
The source-free Gauss-singlet sector satisfies $\hat G^a_\ell\ket\Psi=0$ for all three components.

\subsubsection{Five-state link space}

Following the representation-based construction of Zohar and Burrello~\cite{zohar2015formulation}, we retain the complete $j=0$ and $j=\tfrac12$ representations on every link:
\begin{equation}
\mathcal H_{\mathrm{link}}=\operatorname{span}\{\ket{s},\ket{\mathrm{rr}},\ket{\mathrm{rg}},\ket{\mathrm{gr}},\ket{\mathrm{gg}}\},
\qquad \ket{s}\equiv\ket{0;0,0}.
\label{eq:linkspace}
\end{equation}
The two labels of $\ket{cc'}$ refer to the left dual and right fundamental representation indices, respectively, not to atomic internal states.
Writing $T^a=\sigma^a/2$, the generators vanish on $\ket{s}$ and act in the four-dimensional block as
\begin{equation}
\hat L^a=-T^{a\mathsf T}\otimes\mathbb I_2,
\qquad \hat R^a=\mathbb I_2\otimes T^a,
\qquad
\hat E^2=\hat{\mathbf L}^2=\hat{\mathbf R}^2=\frac34\sum_{c,c'}\ket{cc'}\bra{cc'}.
\label{eq:link_generators}
\end{equation}
With this convention the transporter is
\begin{equation}
\hat U=\frac1{\sqrt2}
\begin{pmatrix}
\ket{\mathrm{rr}}\bra{s}+\ket{s}\bra{\mathrm{gg}} &
\ket{\mathrm{rg}}\bra{s}-\ket{s}\bra{\mathrm{gr}}\\
\ket{\mathrm{gr}}\bra{s}-\ket{s}\bra{\mathrm{rg}} &
\ket{\mathrm{gg}}\bra{s}+\ket{s}\bra{\mathrm{rr}}
\end{pmatrix}.
\label{eq:U_five}
\end{equation}
This is Eq.~(75) of Ref.~\cite{zohar2015formulation}, written with ket--bra operators instead of auxiliary creation and annihilation operators.
The factor $1/\sqrt2$ follows from the SU(2) Clebsch--Gordan coefficients for transitions between $j=0$ and $j=\tfrac12$.

Direct substitution gives
\begin{align}
[\hat L^a,\hat U^{cc'}]&=-\sum_dT^a_{cd}\hat U^{dc'},&
[\hat R^a,\hat U^{cc'}]&=\sum_d\hat U^{cd}T^a_{dc'},\nonumber\\
[\hat L^a,\hat L^b]&=i\epsilon^{abc}\hat L^c,&
[\hat R^a,\hat R^b]&=i\epsilon^{abc}\hat R^c,\qquad[\hat L^a,\hat R^b]=0.
\label{eq:algebra}
\end{align}
These identities ensure exact gauge covariance of the truncated Hamiltonian.
The truncation changes the link Hilbert space while retaining the noncommuting SU(2) generators.
The projected transporter is not a unitary matrix of operators, as it omits transitions to $j\ge1$; its transformation law, rather than unitarity, ensures the gauge invariance required here.
A $j=\tfrac12$ link has electric energy $3g^2/8$, and a $j=0$ link has zero electric energy.

\subsection{The selected gauge sector and its hadronic dynamics}
\label{sm:lgt:sector}

The selected Gauss sector retains the two-quark color-singlet channel of SU(2), supporting baryons and antibaryons alongside mesons.
We first establish this hadron structure and then compare its dynamics with the source-free SU(2) sector and a two-flavor U(1) model with matched local conversion parameters.

\subsubsection{Local constraints and boundary conditions}

We select the highest-weight sector
\begin{equation}
\hat{\mathbf G}_\ell^2=2,\qquad \hat G^z_\ell=1
\label{eq:sector_charge}
\end{equation}
at every matter site.
For $j\le\tfrac12$, at most three spin-$\tfrac12$ contributions meet at a site: one matter charge and two link ends.
The condition $G^z_\ell=1$ requires exactly two such contributions, both with projection $+\tfrac12$.
Their product is the highest-weight state of total spin one, and hence also satisfies $\mathbf G_\ell^2=2$.
This fixes singly occupied matter to $\ket{\mathrm r}$ and each nonzero link to
\begin{equation}
\ket{\mathbf1}\equiv\ket{\mathrm{gr}},\qquad
\ket{\mathbf0}\equiv\ket{s}.
\label{eq:sector_flux}
\end{equation}
Although the left representation index of $\ket{\mathrm{gr}}$ is $\mathrm g$, its left electric-field projection is $+\tfrac12$ because the left index transforms in the dual representation [Eq.~\eqref{eq:link_generators}].
Both electric-field ends therefore have projection $+\tfrac12$.

Let $\lambda_\ell=1$ for $\ket{\mathbf1}$ and $\lambda_\ell=0$ for $\ket{\mathbf0}$.
The allowed local states satisfy
\begin{equation}
\lambda_{\ell-1}+\lambda_\ell+\mathbf1_{n^\psi_\ell=1}=2,
\qquad\lambda_{-1}=\lambda_{L-1}=1.
\label{eq:gauss}
\end{equation}
Here $\mathbf1_{n^\psi_\ell=1}$ equals one for a singly occupied matter site and zero otherwise.
The last two conditions fix the exterior boundary link ends.
A singly occupied site has one adjacent zero-flux link; an empty or doubly occupied matter site has two $j=\tfrac12$ links.
As neighboring sites share the same link, zero-flux links join pairs of singly occupied sites and cannot touch one another.
This gives a complete local characterization of the selected Hilbert space.

The sector has nonzero Gauss charge, with the spin-one (adjoint) representation at each site.
An explicit static spin-one source can be coupled to each dynamical charge to form a total Gauss singlet.
For example, a dynamical multiplet with internal label $\nu$ combines with a source $q$ as
\begin{equation}
\ket{\nu;\mathrm{singlet}}=
\frac1{\sqrt3}\sum_{M=-1}^{1}(-1)^{1-M}
\ket{\nu;1,M}\ket{q;1,-M}.
\label{eq:background_singlet}
\end{equation}
Because the Hamiltonian commutes with all three Gauss generators, it acts identically on the magnetic components of a given multiplet.
This construction relates the selected sector to a gauge theory with static adjoint sources.
The source-free sector $\mathbf G_\ell^2=0$, used for the comparison below, has a different local constraint.

\subsubsection{Hadrons and conserved net baryon number}
\label{sm:lgt:crystals}
\label{sm:lgt:observables}

For the staggered-fermion convention of the main text, even sites host quarks and holes in the filled odd sublattice represent antiquarks.
The bare vacuum has occupations $(0,2,0,2,\ldots)$ and $j=\tfrac12$ on every link.
All hadrons in the selected sector have compact bound-state configurations: a meson occupies neighboring singly occupied particle and antiparticle sites joined by a $j=0$ link, whereas a baryon or antibaryon occupies one site.
The two possible bond parities define $M$ and $M'$.
The bare vacuum and these compact configurations are basis states; hopping builds their quantum superpositions.

The baryon creation operator on a particle site is
\begin{equation}
\hat b^\dagger_\ell=\hat\psi^\dagger_{\ell\mathrm r}\hat\psi^\dagger_{\ell\mathrm g},
\qquad [\hat G^a_k,\hat b^\dagger_\ell]=0.
\label{eq:baryon_operator}
\end{equation}
It creates the antisymmetric two-color singlet; removing the corresponding pair from a filled antiparticle site creates an antibaryon \cite{raychowdhury2020loop,atas2021hadrons}.
Mesonic excitations are generated by gauge-invariant matter hopping, which rearranges both the matter and the connecting link.
Thus the meson includes the change in electric flux, rather than just the two matter occupations.

For even $L$, the net baryon number and the density used in the main text are
\begin{equation}
\hat B=\frac12\sum_\ell(\hat n^\psi_\ell-1),\qquad
\rho_B=\frac{2B}{L},\qquad [\hat H_{\mathrm{LGT}},\hat B]=0.
\label{eq:baryon_number}
\end{equation}
The quark and antiquark excitation counts are $N_q=\sum_{\ell\,\mathrm{even}}n^\psi_\ell$ and $N_{\bar q}=\sum_{\ell\,\mathrm{odd}}(2-n^\psi_\ell)$, so that $B=(N_q-N_{\bar q})/2=N_B-N_{\bar B}$.
The relation $[\hat B,\hat b^\dagger_\ell]=\hat b^\dagger_\ell$, together with Eq.~\eqref{eq:baryon_operator}, shows that a baryon adds one unit of net baryon number while preserving every local Gauss charge.
Its creation on an empty particle site leaves the adjacent links unchanged: empty and doubly occupied matter states have the same flux requirement in Eq.~\eqref{eq:gauss}.
The selected Gauss sector therefore supports initial states with different conserved net baryon numbers at fixed boundary fluxes.
Antibaryons carry $B=-1$ and mesons $B=0$.
This independence of gauge charge and net baryon number~\cite{atas2021hadrons} underlies the finite-density dynamics studied in the main text.

For U(1) matter fields carrying the same gauge charge, the particle--antiparticle imbalance is tied to the net gauge charge.
For equal boundary electric fields and no external charge, the sum of the U(1) Gauss constraints fixes the total particle--antiparticle imbalance to zero.
A pair of such particles carries twice the elementary U(1) charge, including when the two particles belong to different flavors.
In SU(2), the antisymmetric combination of two fundamental colors instead forms a gauge-neutral baryon with $B=1$.
Color-singlet baryons, meson--baryon interconversion and an independent net baryon number constitute the non-Abelian hadron content retained by the selected sector.

\subsubsection{Non-Abelian features of hadron dynamics}
\label{sm:lgt:twolevel}

Consider a particle--antiparticle bond with exterior flux $j=\tfrac12$.
Its compact meson $\ket M$, baryon--antibaryon pair $\ket{B\bar B}$ and bare vacuum $\ket\Omega$ have matter occupations $(1,1)$, $(2,0)$ and $(0,2)$, respectively.
Relative to the meson energy, their detunings are
\begin{equation}
\varepsilon_B=\frac38g^2+2m,\qquad
\varepsilon_V=\frac38g^2-2m.
\label{eq:detunings}
\end{equation}
In the ordered basis $\{\ket M,\ket{B\bar B},\ket\Omega\}$, the two-site Hamiltonian is
\begin{equation}
\hat h_{\mathrm{cell}}=
\begin{pmatrix}
0&-\kappa/\sqrt2&-\kappa/\sqrt2\\
-\kappa/\sqrt2&\varepsilon_B&0\\
-\kappa/\sqrt2&0&\varepsilon_V
\end{pmatrix}.
\label{eq:cell_hamiltonian}
\end{equation}
The two resonances lie at $m=\mp3g^2/16$.

The same two-site matrix also occurs in the source-free Gauss-singlet sector when the two exterior link representations are fixed to $j=\tfrac12$.
In that sector the matter colors and link-end indices are coupled into local singlets instead of highest-weight states.
An empty or doubly occupied matter site with two half-integer link ends has tensor $\delta_{ac}/\sqrt2$, where $a$ is the incoming fundamental index and $c$ the outgoing dual index.
A singly occupied site couples its matter color to the outgoing dual index through $\delta/\sqrt2$, or to the incoming fundamental index through $\epsilon/\sqrt2$, with $\epsilon=i\sigma^y$.
Contracting these normalized local tensors with Eq.~\eqref{eq:U_five} gives hopping magnitude $\kappa/\sqrt2$ for both conversion channels.
A change of the meson basis phase makes the signs identical to Eq.~\eqref{eq:cell_hamiltonian}.
Thus the local resonance positions and population oscillations are exactly the same for this matched two-site comparison.

On a longer chain, conversion processes on both bond parities couple neighboring cells and allow meson regeneration on shifted bonds.
The Gauss-singlet sector additionally allows empty or doubly occupied matter sites with two $j=0$ links, whereas Eq.~\eqref{eq:gauss} excludes these configurations in the selected sector.
The two sectors therefore share the local conversion matrix but have different many-body configuration spaces.
To compare their spatial dynamics, we construct both Hamiltonians from the five-state link operators on an open $L=16$ chain, with identical $m$, $g^2$, $\kappa$ and exterior $j=\tfrac12$ representations.
The bare-vacuum reference has occupations $(0,2,0,2,\ldots)$ and $j=\tfrac12$ on every link in both sectors.

To compare gauge structures with two fermionic matter components per site, we consider two equal-mass flavors carrying the same U(1) charge.
We use spin-one links, the smallest symmetric quantum-link representation for which the standard quadratic electric energy $E^2$ is nonconstant.
The Hamiltonian reads
\begin{equation}
\hat H_{\mathrm{U(1)}}=-w\sum_{\ell}\sum_{f=1}^{2}
\left(\hat c^\dagger_{\ell f}\frac{\hat S^+_\ell}{\sqrt2}\hat c_{\ell+1,f}+\mathrm{h.c.}\right)
+m\sum_\ell(-1)^\ell\hat n^{\mathrm U}_\ell
+K\sum_\ell\hat E_\ell^2,
\qquad \hat E_\ell=\hat S^z_\ell .
\label{eq:u1_reference}
\end{equation}
Here $\hat c_{\ell f}$ annihilates a fermion of flavor $f$, $\hat n^{\mathrm U}_\ell=\sum_f\hat c^\dagger_{\ell f}\hat c_{\ell f}$, and the Gauss constraint is $E_\ell-E_{\ell-1}=n^{\mathrm U}_\ell-[1-(-1)^\ell]$.
The exterior and bare-vacuum fields are $E=-1$.
We set $K=3g^2/8$ and $w=\kappa/2$, so that the meson created by flavor-summed hopping couples to both the vacuum and the doubly occupied particle--empty antiparticle configuration with magnitude $\kappa/\sqrt2$, matching Eq.~\eqref{eq:cell_hamiltonian}.
The two on-site matter pairs in the latter configuration carry opposite U(1) charges, in contrast to the color-singlet baryon and antibaryon in SU(2).

For all three models, we choose $K/\kappa\simeq2.06$ and $m=-K/2$, satisfying the conversion resonance $2m+K=0$ with vacuum detuning $K-2m=2K\simeq4.11\kappa$.
One compact meson is placed on the central even--odd cell $(6,7)$, with all other sites in the bare-vacuum pattern.
In U(1), the initial meson is obtained by applying the flavor-summed hopping operator to the vacuum and normalizing the state.
All three initial states have zero particle--antiparticle imbalance.
To isolate the response to this excitation, we subtract a separate evolution from the bare vacuum:
\begin{equation}
\delta n_m(\ell,t)=\langle\hat n_{m,\ell}\rangle_{M,t}
-\langle\hat n_{m,\ell}\rangle_{\Omega,t},
\qquad
\hat n_{m,\ell}=
\begin{cases}
\hat n_\ell,&\ell\ \mathrm{even},\\
2-\hat n_\ell,&\ell\ \mathrm{odd}.
\end{cases}
\label{eq:matter_response}
\end{equation}
Here $\hat n_\ell$ denotes the total matter occupation in the corresponding model.
We evolve matrix product states with fourth-order time-evolving block decimation~\cite{vidal2004efficient}, using maximum bond dimension $128$ and time step $\kappa\Delta t/\hbar=0.04$, up to $\kappa t/\hbar=40$.

\begin{figure}[htbp]
\centering
\includegraphics[width=\linewidth]{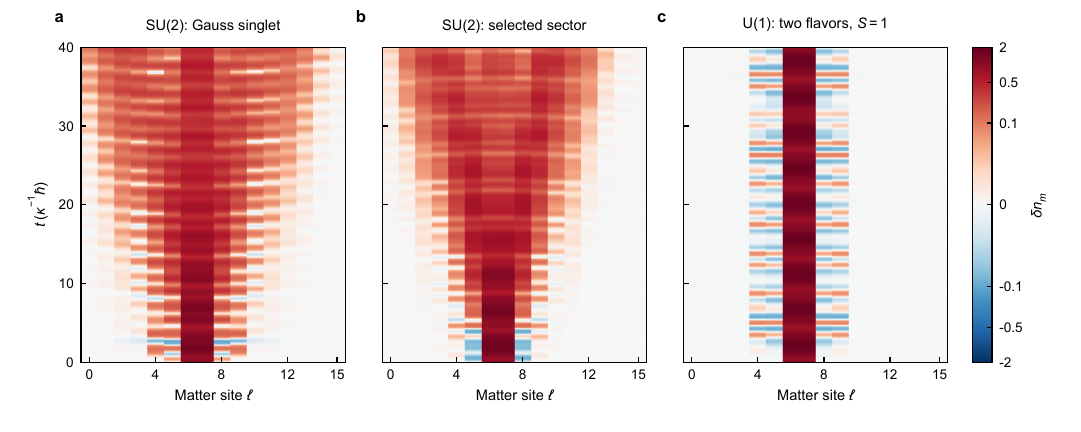}
\caption{\textbf{Single-meson density response in SU(2) and U(1) gauge theories.}
\textbf{a}, Source-free SU(2) Gauss-singlet sector.
\textbf{b}, Selected SU(2) sector $\mathbf G_\ell^2=2$, $G_\ell^z=1$.
\textbf{c}, Two-flavor U(1) model with spin-one links.
The heatmaps show the excess matter density defined in Eq.~\eqref{eq:matter_response} on an open $L=16$ chain, starting from one meson on sites $(6,7)$.
All models have matched local conversion amplitudes, $K/\kappa\simeq2.06$ and $m/\kappa\simeq-1.03$.
The exterior and bare-vacuum links carry $j=\tfrac12$ in SU(2) and $E=-1$ in U(1).
All panels share a symmetric-logarithmic color scale from $-2$ to $2$, linear for $|\delta n_m|\le0.05$.
Negative values indicate a reduced density relative to the separately evolved vacuum reference.}
\label{fig:sector_heatmap}
\end{figure}

Figure~\ref{fig:sector_heatmap} shows an outward-spreading density response in both SU(2) sectors, whereas the U(1) response remains concentrated near the initial meson over the same time interval.
The SU(2) profiles differ in their spreading and oscillation patterns, consistent with the additional configurations available in the Gauss-singlet sector.
Matching the local conversion matrix elements and detunings makes this comparison sensitive to how the conversion processes connect across the chain.
The selected sector thus shares spatially spreading hadron dynamics with the Gauss-singlet theory, while the matched Abelian response remains concentrated near the initial meson.

\FloatBarrier
\subsubsection{Many-body spectral correspondence and the Abelian comparison}
\label{sm:lgt:spectra}

The fixed SU(2) sector and the Gauss-singlet sector share an exact Hamiltonian structure.
Map each fixed-sector configuration to the Gauss-singlet configuration with the same matter occupations and link representations, using the corresponding normalized color tensors and a phase $(-1)^{N_0}$, where $N_0$ counts internal $j=0$ links.
The resulting isometry $W$ satisfies
\begin{equation}
W^\dagger W=\mathbf{1},\qquad
W^\dagger\hat H_{\mathrm{sing}}W=\hat H_{\mathrm{sel}}.
\label{eq:sector_compression}
\end{equation}
The mass and electric energies coincide configuration by configuration, and the hopping matrix elements agree after this phase choice.
Coupling to additional Gauss-singlet configurations determines how this shared structure appears in the many-body spectrum.

\begin{figure}[htbp]
\centering
\includegraphics[width=\linewidth]{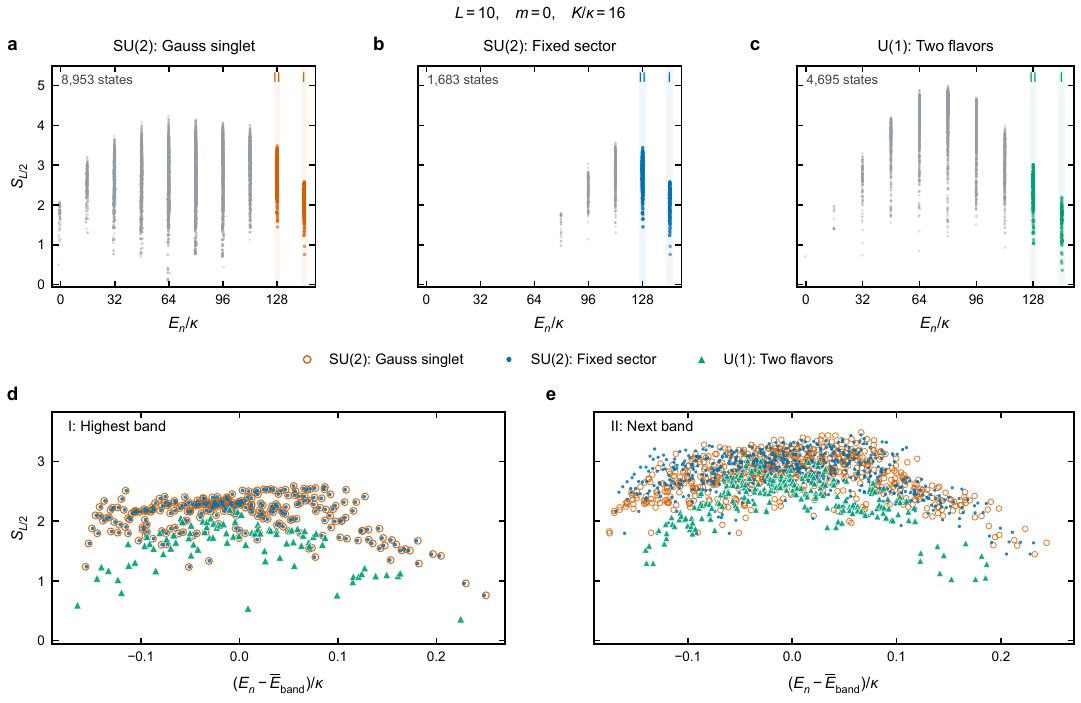}
\caption{\textbf{Many-body spectra and eigenstate entanglement in SU(2) and U(1).}
All models use open $L=10$ chains with total matter occupation $N=L$, $m=0$, $K/\kappa=16$ and the matched local couplings defined in Sec.~\ref{sm:lgt:twolevel}.
The exterior links have $j=\tfrac12$ in SU(2) and $E=-1$ in U(1).
\textbf{a--c}, Complete spectra in the Gauss-singlet SU(2), fixed SU(2) and global-flavor-singlet two-flavor U(1) sectors, respectively.
Each point represents an eigenstate with energy $E_n$ and half-chain entropy $S_{L/2}=-\operatorname{Tr}(\rho_A\ln\rho_A)$, where $\rho_A$ is obtained by tracing out the right half of the chain in the orthonormal dressed-site basis.
The highest two bands are highlighted as I and II.
\textbf{d,e}, Comparison of bands I and II, containing respectively 252 and 630 states in each SU(2) sector, and 89 and 235 states in U(1).
Energies are measured from the arithmetic mean $\overline E_{\mathrm{band}}$ of the corresponding band in each model, retaining the common energy unit $\kappa$.
Orange open circles, blue points and green triangles denote the Gauss-singlet SU(2), fixed SU(2) and U(1) sectors.}
\label{fig:sector_spectra}
\end{figure}

The two sectors develop common many-body bands in the strong-coupling regime $m=0$ and $K\gg\kappa$, where $K=3g^2/8$.
The highest band is built from configurations with $j=\tfrac12$ on every internal link and matter occupation $n_\ell=0$ or $2$ at each site.
These configurations describe arrangements of vacuum, baryons and antibaryons with the same bare electric energy $(L-1)K$.
Virtual hopping through compact-meson configurations couples neighboring $\ket{20}$ and $\ket{02}$ states.
In both SU(2) sectors, this gives the same effective Hamiltonian to second order,
{
\begin{equation}
\hat H_{\mathrm I}^{\mathrm{eff}}
=(L-1)K\,\mathbf{1}
+\frac{\kappa^2}{2K}\sum_{\ell=1}^{L-1}
\bigl(\ket{20}+\ket{02}\bigr)_{\ell,\ell+1}
\bigl(\bra{20}+\bra{02}\bigr)_{\ell,\ell+1}
+O\!\left(\frac{\kappa^4}{K^3}\right).
\label{eq:shared_many_body_band}
\end{equation}
}
Here the two-site kets specify the matter occupations within this band.
The shared pair exchange produces closely matching energy levels and eigenstate entanglement profiles, as shown in Fig.~\ref{fig:sector_spectra}d.

The next band contains one compact meson among the empty and doubly occupied sites.
Its leading effective Hamiltonians are related by a staggered phase choice for the meson position, together with a uniform energy shift and a correction on the end bonds.
To compare the internal band structure, Fig.~\ref{fig:sector_spectra}e plots energies relative to each model's band centre.
The two SU(2) sectors again show similar energy and entanglement distributions, with residual differences associated with the boundaries and higher-order virtual processes.

The two-flavor, spin-one U(1) model of Eq.~\eqref{eq:u1_reference} provides a comparison with the same local conversion amplitude and electric-energy scale.
Its on-site matter pairs carry gauge charge, so their allowed arrangements are tied to the electric-field orientation.
The complete spectra in Fig.~\ref{fig:sector_spectra}a--c place the two highlighted bands within the different many-body Hilbert spaces.
Within these bands, the U(1) model has distinct level distributions and generally lower half-chain entanglement than the two SU(2) sectors (Fig.~\ref{fig:sector_spectra}d,e).
This comparison identifies a common strong-coupling many-body hadronic structure in the two SU(2) sectors, distinct from that of the matched Abelian model.

\FloatBarrier

\subsection{Atomic encoding, exact projection and validity conditions}
\label{sm:lgt:mapping}
\label{sm:lgt:verification}

\subsubsection{One-to-one state map}

Table~\ref{tab:basis} maps the local atomic states to the matter occupation and the two neighboring link states.
The two atomic species distinguish the location of the adjacent zero-flux link, rather than representing two freely variable matter colors.

\begin{table}[htb]
\caption{\textbf{Local atomic encoding.} Each row gives the incoming link, matter state and outgoing link in the selected Gauss-charge sector. The flux labels are defined in Eq.~\eqref{eq:sector_flux}. A single atom on an even (odd) site represents a quark (antiquark); complete mesons are defined by the shared-link constraint.}
\label{tab:basis}
\begin{ruledtabular}
\begin{tabular}{ccc}
Atomic state & $(\text{left link},\text{matter},\text{right link})$ & even / odd site\\
\hline
$\ket0$ & $\ket{\mathbf1,0,\mathbf1}$ & $\Omega_h$ / $\bar B$\\
$\ket\alpha$ & $\ket{\mathbf1,\mathrm r,\mathbf0}$ & quark / antiquark\\
$\ket\beta$ & $\ket{\mathbf0,\mathrm r,\mathbf1}$ & quark / antiquark\\
$\ket{\alpha\beta}$ & $\ket{\mathbf1,\mathrm{rg},\mathbf1}$ & $B$ / $\Omega_d$\\
\end{tabular}
\end{ruledtabular}
\end{table}

A shared zero-flux link requires $\alpha$ on its left and $\beta$ on its right.
The atomic basis therefore consists of all tilings by directed $(\alpha,\beta)$ dimers, holes and $\alpha \beta$ doublons, with no unpaired single atoms at the boundaries.
Every such tiling determines one gauge-theory basis state, and Eq.~\eqref{eq:gauss} reconstructs a unique tiling from every gauge-theory basis state in the selected sector.
Let $\mathcal B_G$ denote this orthonormal gauge-theory basis, with fermionic creation operators ordered as $(0,\mathrm r),(0,\mathrm g),(1,\mathrm r),(1,\mathrm g),\ldots$.
For $x\in\mathcal B_G$, write $a(x)$ for its corresponding atomic tiling.

The isometry implementing the encoding is
\begin{equation}
\mathcal V=\sum_{x\in\mathcal B_G}\ket{a(x)}_{\mathrm{at}}\,{}_{\mathrm G}\!\bra{x},
\qquad
\mathcal V^\dagger\mathcal V=\mathbb I_G,\qquad
\mathcal V\mathcal V^\dagger=\mathcal P_{\mathrm{at}},
\label{eq:encoding_isometry}
\end{equation}
where $\mathbb I_G$ is the identity on the selected gauge sector and $\mathcal P_{\mathrm{at}}$ projects onto the encoded atomic space.
The two identities follow from the orthonormality and bijection of the bases.
The dimension obeys $d_L=2d_{L-1}+d_{L-2}$, with $d_0=1$ and $d_1=2$.
At fixed atom number $N=2q$ it is
\begin{equation}
d_{L,2q}=\sum_{k=0}^{\lfloor L/2\rfloor}
\binom{L-k}{k}\binom{L-2k}{q-k},
\label{eq:sector_dimension}
\end{equation}
where binomial coefficients outside their allowed range are zero.
Here $k$ counts dimers and $q-k$ counts doublons.
For $L=N=20$, the dimension is $8\,097\,453$.

The map preserves the total occupation, $n^\psi_\ell=n_{\ell\alpha}+n_{\ell\beta}$, and hence $B=(N-L)/2$ and $\rho_B=N/L-1$.
The net baryon density is therefore controlled by atom number relative to the encoded vacuum.
The ideal staggered polarization is
\begin{equation}
\hat\Sigma=\frac1L\sum_\ell(-1)^\ell(\hat n_{\ell\alpha}+\hat n_{\ell\beta}).
\label{eq:sigma}
\end{equation}
It equals $+1$, $0$ and $-1$ in the fully baryonic, N\'eel and bare-vacuum configurations, respectively.
The local matter excitation is $n_m=n^\psi_\ell$ on even sites and $n_m=2-n^\psi_\ell$ on odd sites, and the reconstructed electric Casimir is $E^2_\ell=3\lambda_\ell/4$.

\subsubsection{Hopping matrix elements and fermionic signs}
\label{sm:lgt:jw}

We evaluate the hopping using the color-resolved matrix elements of the SU(2) transporter in Eq.~\eqref{eq:U_five}.
Between the allowed link states $\ket{\mathbf0}=\ket{s}$ and $\ket{\mathbf1}=\ket{\mathrm{gr}}$, its nonzero matrix elements are
\begin{equation}
\bra{\mathbf0}\hat U_\ell^{\mathrm{rg}}\ket{\mathbf1}=-\frac1{\sqrt2},
\qquad
\bra{\mathbf1}\hat U_\ell^{\mathrm{gr}}\ket{\mathbf0}=\frac1{\sqrt2}.
\label{eq:sector_link_matrix_elements}
\end{equation}
In the matter--gauge hopping, these two color channels connect the $j=0$ and $j=\tfrac12$ link representations.
Let $\mathcal P_G$ project onto the joint matter--link Hilbert space satisfying Eq.~\eqref{eq:sector_charge} and the boundary conditions in Eq.~\eqref{eq:gauss}.
The hopping within this sector is
\begin{equation}
\begin{split}
\hat H_{\mathrm{hop}}^{\mathrm{sector}}
&=\mathcal P_G\hat H_{\mathrm{hop}}\mathcal P_G\\
&=-\kappa\sum_{\ell,c,c'}\mathcal P_G
\left(\hat\psi^\dagger_{\ell c}\hat U_\ell^{cc'}\hat\psi_{\ell+1,c'}
+\mathrm{h.c.}\right)\mathcal P_G .
\end{split}
\label{eq:sector}
\end{equation}
Each allowed transition changes the matter and link states together while preserving the local Gauss charges.
The Jordan--Wigner sign for the $(c,c')=(\mathrm r,\mathrm g)$ channel is $(-1)^{n_{\ell\mathrm g}+n_{\ell+1,\mathrm r}}$.
For every allowed transition in this channel, $n_{\ell\mathrm g}+n_{\ell+1,\mathrm r}=1$, so the fermionic sign cancels the negative transporter matrix element in Eq.~\eqref{eq:sector_link_matrix_elements}.
The $(\mathrm g,\mathrm r)$ channel connects adjacent fermionic modes and has no intermediate string.
Both channels therefore have matrix element $-\kappa/\sqrt2$.

A nearest-neighbor hop changes the matter occupation at both ends of one bond while leaving its exterior links fixed.
Equation~\eqref{eq:gauss} then requires both exterior links to be $\ket{\mathbf1}$, with the central link changing between $\ket{\mathbf1}$ and $\ket{\mathbf0}$.
Particle-number conservation leaves exactly the three local configurations $(n^\psi_\ell,n^\psi_{\ell+1},\lambda_\ell)=(2,0,1),(1,1,0),(0,2,1)$.
Thus all nonzero projected atomic hops belong to
\begin{equation}
\ket{\alpha\beta,0}\ \longleftrightarrow\ \ket{\alpha,\beta}\
\longleftrightarrow\ \ket{0,\alpha\beta}.
\label{eq:processes}
\end{equation}
Each allowed atomic hop moves one atom of a given species between occupations zero and one and has matrix element $-J$, with no same-species ($\alpha \alpha$ or $\beta \beta$) bosonic enhancement.
The matrix elements therefore agree for every encoded pair of states when $\kappa=\sqrt2J$.
Writing $\hat H_{\mathrm{BH}}^{\mathrm{hop}}$ for the tunneling part of Eq.~\eqref{eq:bhm}, this establishes the operator identity
\begin{equation}
\mathcal V^\dagger\mathcal P_{\mathrm{at}}\hat H_{\mathrm{BH}}^{\mathrm{hop}}
\mathcal P_{\mathrm{at}}\mathcal V
=\hat H_{\mathrm{hop}}^{\mathrm{sector}} .
\label{eq:hopping_identity}
\end{equation}

\subsubsection{Diagonal energies and the projected Hamiltonian identity}

The two-component Bose--Hubbard Hamiltonian is defined in Eq.~\eqref{eq:bhm}.
The species-dependent tilt gives energy $-\ell\Delta$ to an $\alpha$ atom and $+\ell\Delta$ to a $\beta$ atom; the staggered potential adds $\delta$ per atom on an even site.
For an even--odd bond, the three configurations in Eq.~\eqref{eq:processes} have energies
\begin{equation}
E_{\alpha\beta,0}=V+2\delta,\qquad
E_{\alpha,\beta}=\Delta+\delta,\qquad
E_{0,\alpha\beta}=V.
\label{eq:energies}
\end{equation}
The corresponding conversion costs are
\begin{equation}
\varepsilon_B=V-\Delta+\delta,\qquad
\varepsilon_V=V-\Delta-\delta.
\label{eq:costs}
\end{equation}
Combining these expressions with Eq.~\eqref{eq:detunings} gives
\begin{equation}
\kappa=\sqrt2J,\qquad m=\frac\delta2,\qquad
g^2=\frac83(V-\Delta).
\label{eq:dictionary}
\end{equation}

The equality of diagonal energies extends to every encoded configuration.
Let $N_d$ and $N_D$ count dimers and doublons, and let $N_{\mathrm{even}}=\sum_{\ell\,\mathrm{even}}n_\ell$ count atoms on the even sublattice.
Each dimer contributes $\Delta$ to the tilt energy, while a doublon has zero net tilt energy; $N/2=N_d+N_D$ and $\sum_\ell\lambda_\ell=L-1-N_d$.
Also, $\delta\sum_{\ell\,\mathrm{even}}n_\ell=(\delta/2)\sum_\ell(-1)^\ell n_\ell+\delta N/2$.
For any encoded basis state $x$, its atomic diagonal energy can consequently be written as
\begin{equation}
\begin{split}
E_{\mathrm{at}}(x)
&=\Delta N_d+VN_D+\delta N_{\mathrm{even}}\\
&=\frac{\delta}{2}\sum_\ell(-1)^\ell n_\ell
+(V-\Delta)\sum_{\ell=0}^{L-2}\lambda_\ell
+\frac{V+\delta}{2}N-(V-\Delta)(L-1).
\end{split}
\label{eq:diagonal_energy_identity}
\end{equation}
The first two terms on the second line are the target mass and electric energies under Eq.~\eqref{eq:dictionary}.
Combining this equality on every basis state with Eq.~\eqref{eq:hopping_identity} gives
\begin{equation}
\mathcal V^\dagger\mathcal P_{\mathrm{at}}\hat H_{\mathrm{BH}}\mathcal P_{\mathrm{at}}\mathcal V
=\hat H_{\mathrm{LGT}}\big|_{(\mathbf G^2,G^z)=(2,1)}
+\frac{V+\delta}{2}\hat N-(V-\Delta)(L-1),
\label{eq:projected_identity}
\end{equation}
At fixed $N$, the last two terms are a common energy offset.
Equations~\eqref{eq:encoding_isometry} and \eqref{eq:projected_identity} establish unitary equivalence between the target Hamiltonian and the atomic Hamiltonian restricted to the encoded space, up to this offset.
Their time evolutions are therefore related by $\mathcal V$ and an overall phase.
The encoding is independent of the Hamiltonian parameters, so the same identity holds at each time during a ramp with the corresponding parameter dictionary.
The intra-species interaction $U$ drops out because each species has occupation at most one in this subspace.
Its role is to control transitions outside the encoding, addressed next.

\subsubsection{Parameter conditions and suppression of leakage}
\label{sm:lgt:validity}

A convenient regime for resonant gauge dynamics is $V,\Delta\gg J$, with the interaction--gradient mismatch and staggered offset setting the target conversion energies $\varepsilon_{B,V}=V-\Delta\pm\delta$.
The desired meson--baryon or meson--vacuum conversion becomes resonant when the corresponding $\varepsilon_{B,V}$ approaches zero.
For repulsive interactions, the single-hop leakage channels from the encoded basis are off resonant under the following conditions:
\begin{equation}
\begin{aligned}
|\Delta\pm\delta|,\quad |V+\Delta\pm\delta|&\gg J,\\
|U-V-\Delta\pm\delta|,\quad |U+V+\Delta\pm\delta|&\gg\sqrt2J,\\
|U-\Delta\pm\delta|,\quad |U+\Delta\pm\delta|&\gg\sqrt2J.
\end{aligned}
\label{eq:parameter_conditions}
\end{equation}
Both signs of $\delta$ are included, accounting for the two bond parities.
The first line suppresses single-atom motion that breaks an encoded dimer, and conversion to or from the reversed pair $(\beta,\alpha)$.
The second line concerns an $\alpha \beta$ doublon next to a single atom: tunneling can create either a same-species ($\alpha \alpha$ or $\beta \beta$) doublon or a three-atom site and a hole.
The third line suppresses atom transfer between adjacent $\alpha \beta$ doublons, producing occupations $(3,1)$ or $(1,3)$.
The factors $\sqrt2$ are the bosonic enhancement of hopping into an already occupied component.

In the near-resonant window $|V-\Delta|,|\delta|=O(J)$, Eq.~\eqref{eq:parameter_conditions} reduces to
\begin{equation}
V\simeq\Delta\gg J,\qquad
|U-V|\gg\sqrt2J,\qquad |U-2V|\gg\sqrt2J.
\label{eq:near_resonant_conditions}
\end{equation}
The two interaction differences separate the neighboring-doublon resonance and the resonance for producing a same-species doublon, respectively.
For a mass ramp with varying $\delta$, Eq.~\eqref{eq:parameter_conditions} gives the detunings throughout the scan.

These conditions follow directly from the occupation-dependent hopping energies.
Write $\hat H_{\mathrm{BH}}=\hat H_0+\hat T$, with $\hat H_0$ diagonal in the atomic occupation basis and $\hat T$ the tunneling term, and let $\mathcal Q_{\mathrm{at}}=1-\mathcal P_{\mathrm{at}}$.
For a hop of species $s$ from site $\ell$ to site $\ell+1$, the exact unperturbed energy difference is
\begin{equation}
\begin{split}
\Delta E_s={}&U(n_{\ell+1,s}-n_{\ell,s}+1)
+V(n_{\ell+1,\bar s}-n_{\ell,\bar s})\\
&+s_s\Delta+\delta(\chi_{\ell+1}-\chi_\ell),
\end{split}
\label{eq:leakage_denominator}
\end{equation}
where $s_\alpha=-1$, $s_\beta=+1$, $\bar s$ denotes the other species, and $\chi_\ell=1$ on even sites and zero on odd sites.
The hopping magnitude is $J\sqrt{n_{\ell s}(n_{\ell+1,s}+1)}$.
Applying this expression to the encoded occupations gives the channels in Eq.~\eqref{eq:parameter_conditions}.
For fixed parameters, perturbative elimination of these off-resonant channels gives~\cite{bravyi2011schrieffer}
$\hat H_{\mathrm{eff}}=\mathcal P_{\mathrm{at}}\hat H_{\mathrm{BH}}\mathcal P_{\mathrm{at}}+\delta\hat H^{(2)}+\cdots$.
The second-order correction has matrix elements
\begin{equation}
(\delta H^{(2)})_{ij}=\frac12\sum_{q\in\mathcal Q_{\mathrm{at}}}
T_{iq}T_{qj}\left(\frac1{E_i-E_q}+\frac1{E_j-E_q}\right),
\label{eq:second_order_correction}
\end{equation}
Here $i,j$ label encoded occupation states and $q$ labels states outside the encoding, with $E_x=\bra{x}\hat H_0\ket{x}$ and $T_{xy}=\bra{x}\hat T\ket{y}$.
Besides suppressing real leakage, the energy separation controls virtual shifts and induced couplings; their accumulated effect depends on the evolution time.

The adjacent-doublon channel is particularly relevant to the experimental parameters.
For example, $\ket{\alpha\beta,\alpha\beta}\to\ket{\beta,\alpha\alpha\beta}$ has hopping magnitude $\sqrt2J$.
For the two downhill channels, the energy denominators are
\begin{equation}
d_\pm=U-\Delta\pm\delta=(U-V)+\varepsilon_{B,V}.
\label{eq:adjacent_doublon_denominator}
\end{equation}
Thus near the selected conversion resonance, $|U-V|\gg\sqrt2J$ separates the adjacent-doublon leakage process from the desired conversion.

At $U\simeq V$, the source of the potentially resonant three-atom transitions can be expressed directly in terms of neighboring-doublon occupations.
Define
\begin{equation}
\hat D_\ell=\ket{\alpha\beta}_\ell\bra{\alpha\beta},\qquad
A_{DD}(t)=\sum_{\ell=0}^{L-2}\langle\hat D_\ell\hat D_{\ell+1}\rangle_t.
\label{eq:adjacent_doublon_weight}
\end{equation}
Within the encoding, neighboring doublons necessarily occupy opposite sublattices and therefore involve both baryon-type and vacuum-type occupations.
The part of $\mathcal Q_{\mathrm{at}}\hat T\mathcal P_{\mathrm{at}}$ describing the two downhill hops from such pairs is
\begin{equation}
\hat K_{DD}=-J\sum_{\ell=0}^{L-2}
\mathcal Q_{\mathrm{at}}
\left(\hat\alpha^\dagger_{\ell+1}\hat\alpha_\ell+
\hat\beta^\dagger_\ell\hat\beta_{\ell+1}\right)
\hat D_\ell\hat D_{\ell+1}\mathcal P_{\mathrm{at}} .
\label{eq:dd_source_operator}
\end{equation}
Each hop has magnitude $\sqrt2J$ and produces a three-atom configuration that uniquely identifies its encoded parent and hopping bond.
These output Fock states are orthogonal, giving
\begin{equation}
\hat K_{DD}^\dagger\hat K_{DD}
=4J^2\mathcal P_{\mathrm{at}}
\left(\sum_\ell\hat D_\ell\hat D_{\ell+1}\right)\mathcal P_{\mathrm{at}},
\qquad
\|\hat K_{DD}\ket{\phi}\|=2|J|\sqrt{A_{DD}[\phi]}
\label{eq:dd_source_norm}
\end{equation}
for any normalized encoded state $\ket{\phi}$.
This relation quantifies the amplitude coupling that state to the downhill three-atom channels.

The experimental mass and electric-energy offsets provide a second separation of scales within the encoded space.
Define $\hat N_{D,\mathrm e}=\sum_{\ell\,\mathrm{even}}\hat D_\ell$ and $\hat N_{D,\mathrm o}=\sum_{\ell\,\mathrm{odd}}\hat D_\ell$.
Using $N_d=N/2-N_{D,\mathrm e}-N_{D,\mathrm o}$ and $N_{\mathrm{even}}=N/2+N_{D,\mathrm e}-N_{D,\mathrm o}$, the projected Hamiltonian at fixed $N$ becomes
\begin{equation}
\hat H_P=\frac{\Delta+\delta}{2}N
+\varepsilon_B\hat N_{D,\mathrm e}
+\varepsilon_V\hat N_{D,\mathrm o}
+\hat T_{\mathrm e}+\hat T_{\mathrm o},
\label{eq:branch_hamiltonian}
\end{equation}
where $\hat T_p=\hat T_p^++\hat T_p^-$ contains the transitions in Eq.~\eqref{eq:processes} that create or annihilate a doublon on sublattice $p$.
The creation and annihilation parts satisfy $[\hat N_{D,p},\hat T_q^\pm]=\pm\delta_{pq}\hat T_q^\pm$.
Writing $K=V-\Delta=3g^2/8\ge0$, the lower-energy conversion branch has detuning $\varepsilon_a=K-|\delta|$, whereas the other branch has detuning
\begin{equation}
\Lambda=K+|\delta|=\frac{3g^2}{8}+2|m|.
\label{eq:branch_separation}
\end{equation}
The active sublattice $a$ is even for negative mass and odd for positive mass; $s$ denotes the opposite sublattice.
At zero mass either choice can be used.
Thus the relevant distance from the simultaneous-conversion point $m=g=0$ is the off-branch detuning $\Lambda$ relative to $J$.

Let $\mathcal R_a$ project onto encoded configurations with $N_{D,s}=0$.
The ideal N\'eel initial state and states obtained by removing complete mesons both lie in this space.
The leading branch-separated Hamiltonian is $\hat H_a=\hat H_{\mathrm d}+\hat T_a$, where $\hat H_{\mathrm d}=\varepsilon_a\hat N_{D,a}+\Lambda\hat N_{D,s}$ and a scalar energy has been omitted.
Since $\hat T_a$ preserves $N_{D,s}$,
\begin{equation}
[\hat H_a,\hat N_{D,s}]=0,\qquad
\hat K_{DD}\mathcal R_a=0,\qquad
A_{DD}^{(a)}(t)=0 .
\label{eq:branch_invariant_space}
\end{equation}
The active-branch dynamics therefore supports meson conversion and regeneration while remaining exactly disconnected from the adjacent-doublon leakage source.
With the non-DD leakage channels detuned as in Eq.~\eqref{eq:parameter_conditions}, $\mathcal R_a$ is also an invariant component of the leading resonant Bose--Hubbard Hamiltonian.
This protection holds throughout that evolution, starting from the specified doublon-free initial states.

The leading correction is set by the off-resonant branch.
For uniform, fixed parameters, define the anti-Hermitian generator
$\hat S_s=(\hat T_s^+-\hat T_s^-)/\Lambda$.
The commutators above give $[\hat S_s,\hat H_{\mathrm d}]=-\hat T_s$, and expansion of the rotated Hamiltonian yields
\begin{equation}
\begin{split}
e^{\hat S_s}(\hat H_{\mathrm d}+\hat T_a+\hat T_s)e^{-\hat S_s}
={}&\hat H_{\mathrm d}+\hat T_a+[\hat S_s,\hat T_a]\\
&+\frac{[\hat T_s^+,\hat T_s^-]}{\Lambda}
+O(J^3/\Lambda^2).
\end{split}
\label{eq:branch_elimination}
\end{equation}
The first correction to local process amplitudes is therefore of order $J^2/\Lambda$.
For a normalized active-branch state $\ket{\varphi_a}\in\mathcal R_a$, the dressed state $e^{-\hat S_s}\ket{\varphi_a}$ has spectator population
\begin{equation}
\langle\hat N_{D,s}\rangle
=\frac{\bra{\varphi_a}\hat T_s^-\hat T_s^+\ket{\varphi_a}}{\Lambda^2}
+O\!\left[(J/\Lambda)^3\right],
\qquad
\sum_\ell\hat D_\ell\hat D_{\ell+1}\le2\hat N_{D,s}.
\label{eq:spectator_population}
\end{equation}
The operator inequality follows because every neighboring-doublon pair contains a doublon on sublattice $s$, and each such site belongs to at most two bonds.
Thus the local neighboring-doublon population enters at order $(J/\Lambda)^2$, and Eq.~\eqref{eq:dd_source_norm} gives a local leakage-source strength of order $J^2/\Lambda$.
These orders apply at fixed $Jt/\hbar$ with the active conversion retained.
Quench-induced spectator-population transients start at the same quadratic order.

The initial-state selection can also be seen directly in the first process that produces neighboring doublons.
Starting from two neighboring mesons, this requires one even-sublattice and one odd-sublattice conversion on the two initially occupied bonds.
On overlapping bonds both ordered creations vanish, while on disjoint bonds they commute.
The two time orderings therefore give the interaction-picture amplitude
\begin{equation}
a_{DD}^{(2)}(t)=-F_{\mathrm e}(t)F_{\mathrm o}(t),\qquad
F_p(t)=\frac1\hbar\int_0^t J(u)
\exp\!\left[\frac{i}{\hbar}\int_0^u\varepsilon_p(v)\,\mathrm dv\right]\mathrm du .
\label{eq:dd_pair_creation}
\end{equation}
Here $\varepsilon_{\mathrm e}=\varepsilon_B$ and $\varepsilon_{\mathrm o}=\varepsilon_V$.
For constant parameters, the leading, fourth-order probability for this local process is
\begin{equation}
P_{DD}^{(4)}(t)=
\frac{16J^4}{\varepsilon_B^2\varepsilon_V^2}
\sin^2\!\left(\frac{\varepsilon_Bt}{2\hbar}\right)
\sin^2\!\left(\frac{\varepsilon_Vt}{2\hbar}\right),
\label{eq:dd_pair_probability}
\end{equation}
with the continuous limit understood when one detuning vanishes.
This expression explicitly suppresses neighboring-doublon formation when either conversion is off resonant.
When both channels are detuned, two off-resonant creations are required; near one resonance, Eq.~\eqref{eq:branch_elimination} retains that conversion to all orders while expanding in $J/\Lambda$.

The residual leakage proceeds through higher-order paths.
For negative mass, one example is the three-hop conversion
$\ket{\alpha,\beta,\alpha,\beta}\to\ket{0,\beta,\alpha\alpha\beta,0}$.
Its bare detuning is $\Delta_3=(U-V)+3K-|\delta|$.
At $\Delta_3=0$, summing the three allowed hopping orders gives
\begin{equation}
\mathcal J_3=-\frac{\sqrt2J^3}{\Lambda}
\left(\frac1{\varepsilon_a}+\frac1{U-2V+2K}\right).
\label{eq:third_order_leakage}
\end{equation}
Two paths pass through neighboring $\alpha \beta$ doublons; the third passes through a same-species doublon.
The expression applies when the intermediate states are detuned, with the resonance position corrected by the second-order shifts in Eq.~\eqref{eq:second_order_correction}.
For $U=V$ and $V\gg K$, its magnitude approaches $\sqrt2J^3/(8K^2)$.
The analogous process occurs on the opposite sublattice at positive mass.
An observation window satisfying $|\mathcal J_3|t/\hbar\ll1$ resolves the order-$J$ hadron dynamics before this higher-order resonance produces an appreciable transition.

During the mass ramps, $K$ and $J$ are fixed and $\Lambda(t)=K+|\delta(t)|$ increases from $K$.
At the selected conversion resonance, $\Lambda=2K$.
Slow variation of the off-branch dressing is controlled by $\hbar|\dot\Lambda|/\Lambda^2\ll1$.
If a ramp crosses the higher-order resonance, its diabatic passage is controlled by the small Landau--Zener parameter $2\pi|\mathcal J_3|^2/(\hbar|\dot\Delta_3|)$.
For the reference quench Hamiltonian at $R$, $K/h=221\,\mathrm{Hz}$ and $|\delta|/h=222.3\,\mathrm{Hz}$, giving $\Lambda/h=443.3\,\mathrm{Hz}$ and $J/\Lambda\simeq0.172$.
For $U=V$, this reference point also has $\Delta_3/h=440.7\,\mathrm{Hz}$.
The same branch selection applies to the ideal finite-density ensembles obtained by removing complete mesons, which also contain no doublons.

Smooth confinement shifts the local conversion detunings by the potential difference between the doublon site and the emptied site.
For a bond joining sites $\ell_a$ and $\ell_s$ on the active and spectator sublattices, the spectator detuning is $\Lambda_\ell=\Lambda+\epsilon_{\ell_s}-\epsilon_{\ell_a}$.
The generator in Eq.~\eqref{eq:branch_elimination} then uses each bond term divided by its local detuning.
Higher-order detunings such as $\Delta_3$ likewise include the trap-energy difference between the final and initial configurations.
Using the $L=20$ confinement profile specified in Sec.~\ref{sm:1d:quench_target}, the smallest spectator detuning at $R$ is approximately $h\times379\,\mathrm{Hz}$, giving $J/\min_\ell|\Lambda_\ell|\simeq0.201$.
The detuning-dependent expressions above also describe the changing suppression along the mass and gauge-coupling scans.
Together, the doublon-free initial states and the separation $\Lambda\gg J$ provide parametric suppression of the resonant three-atom channel even at $U\simeq V$.
Section~\ref{sm:1d} quantifies the finite-parameter corrections and preparation defects in the experimental trajectories.

\section{Additional results and validation of the quantum simulator}
\label{sm:1d}

We first examine the effects of preparation defects and the additional pairing defects generated during evolution.
We then benchmark the measured hadron dynamics against the target Hamiltonian and the full Bose--Hubbard model.

\subsection{Preparation defects and dynamically generated unpaired atoms}
\label{sm:1d:leakage}

\subsubsection{Unpaired atoms as fragment boundaries}
\label{sm:1d:boundaries}

\begin{figure}[htb]
\centering
\includegraphics[width=\textwidth]{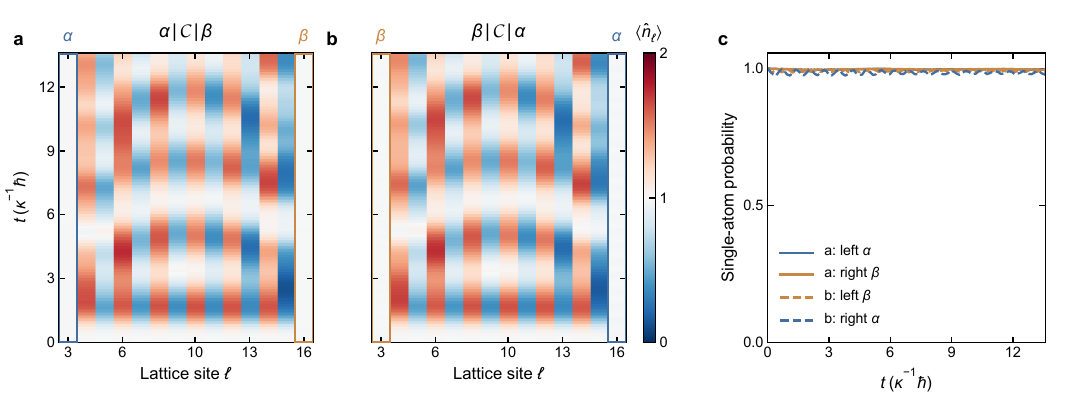}
\caption{\textbf{Stable single-atom boundaries of an encoded fragment.}
Full Bose--Hubbard evolution on a 20-site open chain with $(U,V,\Delta)/h=(8223,1500,1277.7)\,\mathrm{Hz}$, $J/h=76.45\,\mathrm{Hz}$ and $\delta/h=-222.3\,\mathrm{Hz}$.
This interaction-separated control uses the Gaussian profile of Eq.~\eqref{eq:quench_trap} with $w=103\,\mu\mathrm{m}$ and curvature $C\simeq6.62\,\mathrm{Hz}$.
\textbf{a,} $\alpha|\mathcal C|\beta$ boundary orientation.
\textbf{b,} $\beta|\mathcal C|\alpha$ orientation.
The heatmaps show $\langle\hat n_\ell(t)\rangle$ on sites 3--16, including the N\'eel-initialized fragment $\mathcal C=\{4,\ldots,15\}$ and both boundary atoms.
Blue and orange frames mark the initial $\alpha$ and $\beta$ boundary states.
The unshown exterior sites remain part of the calculation.
\textbf{c,} Probability that each framed site retains its initial single-atom state.
Colors match the frames; solid and dashed lines correspond to \textbf{a} and \textbf{b}, respectively.
The dotted line marks unit probability.}
\label{fig:boundary_dressing}
\end{figure}

Figure~\ref{fig:boundary_dressing} illustrates how unpaired atoms delimit an encoded fragment $\mathcal C$ for the two orientations $\alpha|\mathcal C|\beta$ and $\beta|\mathcal C|\alpha$.
In the ideal resonant encoding, these atoms define open boundaries for the surrounding hadronic fragments.
Writing $D\equiv\alpha\beta$, the allowed conversion processes are $(\alpha,\beta)\leftrightarrow(D,0)$ and $(\alpha,\beta)\leftrightarrow(0,D)$.
An unpaired single atom participates in none of these moves: an encoded fragment begins with $0$, $\alpha$ or $D$ and ends with $0$, $\beta$ or $D$.
The resonant Hamiltonian consequently decomposes into independent fragment Hamiltonians and the diagonal energy of the unpaired atoms.
When same-species multiple occupations are energetically separated, including $|U-V|\gg J$ near the conversion resonance, this construction applies to both $\beta|\mathcal C|\alpha$ and $\alpha|\mathcal C|\beta$, where $\mathcal C$ denotes an encoded fragment.

We calculate the two configurations in Fig.~\ref{fig:boundary_dressing}a,b using the full two-component Bose--Hubbard Hamiltonian on 20 sites.
The 12-site fragment $\mathcal C=\{4,\ldots,15\}$ starts in the N\'eel state, with boundary atoms on sites 3 and 16.
All atoms, including those at the boundaries and in the exterior, evolve under the full Hamiltonian.
For $\beta|\mathcal C|\alpha$, we remove the atoms on sites 2 and 17 from the global N\'eel state; the remaining atoms on sites 3 and 16 form the two boundaries.
For $\alpha|\mathcal C|\beta$, we instead reverse the spins on sites 3 and 16.
The benchmark uses the interaction-separated parameters and Gaussian confinement given in the figure caption.
Keeping $V-\Delta$ fixed preserves the target parameters $\kappa$, $m$ and $g^2$, while increasing the separation of unwanted atomic processes.

The heatmaps in Fig.~\ref{fig:boundary_dressing}a,b show the fragment and its two boundary sites; blue and orange frames identify the initial $\alpha$ and $\beta$ atoms.
Figure~\ref{fig:boundary_dressing}c follows their single-atom-state probabilities, $p_b(t)=\langle\hat P_{\sigma_b,\ell_b}\rangle_t$, where $\ell_b=3,16$ and $\sigma_b$ is the corresponding initial internal state.
The projector $\hat P_{\sigma_b,\ell_b}$ selects exactly one atom in state $\sigma_b$ at site $\ell_b$.
All four probabilities remain close to unity, with a minimum of approximately $0.973$ in the calculated trajectories.
The boundary atoms thus retain their single-atom character while the fragment undergoes density oscillations.
Its mean atom number changes by less than $0.05\%$ over 20 ms.
Cutting the two interface bonds in a reference calculation changes the mean meson site fraction by less than $0.015$ for either orientation.
These comparisons establish effective open boundaries for the internal hadron dynamics in this interaction-separated regime.

The calculations use finite, charge-conserving matrix-product states with up to two atoms of each species per site, bond dimension 128 and a $0.005\,\mathrm{ms}$ time step.
Doubling the bond dimension from 64 and halving the time step changes the displayed occupations by less than $0.002$ and boundary-state probabilities by less than $4\times10^{-4}$.
Independent six-site exact evolution agrees within $6\times10^{-4}$; increasing the local cutoff from two to three changes the short-time local densities by less than $8\times10^{-6}$.

For $U\simeq V$, an $\alpha$ on the left or a $\beta$ on the right can hybridize with the nearest dimer through a same-species hop of amplitude $\sqrt2J$ and detuning $U-\Delta\pm\delta$.
The local conversions $\ket{\alpha;\alpha,\beta}\leftrightarrow\ket{0;\alpha\alpha,\beta}$ and $\ket{\alpha,\beta;\beta}\leftrightarrow\ket{\alpha,\beta\beta;0}$ expose a $\beta$ at the new left edge or an $\alpha$ at the new right edge.
At $\delta=\Delta-V$, the leading resonant hopping conserves the two-site species occupations $n_{2j-1,\alpha}+n_{2j,\alpha}$ and $n_{2j,\beta}+n_{2j+1,\beta}$, confining the excess occupation to its boundary link.
This gives a locally dressed boundary in the resonant description.
The full Bose--Hubbard checks at $U=V$ likewise give a boundary-dominated excess population, with smaller finite tails into the fragment interior.

\FloatBarrier
\subsubsection{Dynamically generated defects}
\label{sm:1d:generated_defects}

We quantify defect production during evolution through the increase in unpaired single-atom detections relative to the initial readout.
Directed neighboring $(\alpha,\beta)$ pairs on either bond sublattice and on-site $\alpha\beta$ pairs are first identified in each reconstructed image.
The remaining single detections define
\begin{align}
    \epsilon_{\mathrm{rec}}(t)&=
    \frac{\sum_s N_{\mathrm{unpaired}}^{(s)}(t)}
         {\sum_s N_{\mathrm{det}}^{(s)}(t)},
    \label{eq:reconstructed_defect_fraction}\\
    \Delta\epsilon_{\mathrm{rec}}(t)&=
    \epsilon_{\mathrm{rec}}(t)-\epsilon_{\mathrm{rec}}(t_0),
    \qquad t_0=0.1\,\mathrm{ms},
    \label{eq:generated_defect_increment}
\end{align}
where $s$ labels experimental realizations and both counts use the full reconstructed parent chains before fragment segmentation.
This readout-based observable follows the additional pairing defects generated over the measured evolution window.

\begin{figure}[htbp]
\centering
\includegraphics[width=\textwidth]{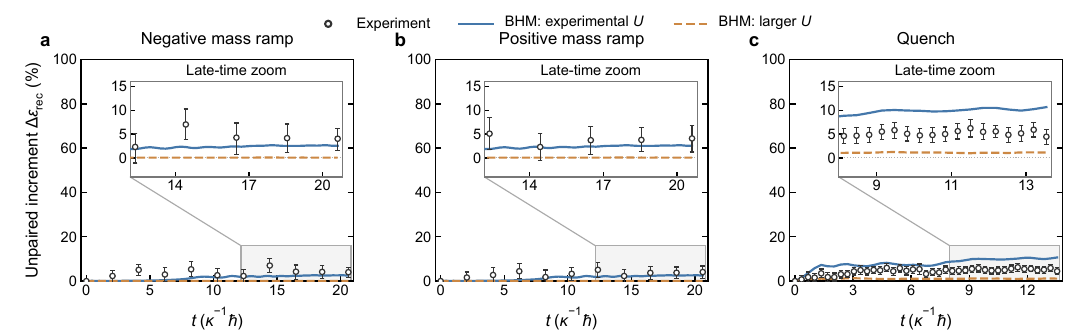}
\caption{\textbf{Dynamically generated unpaired atoms.}
Increment $\Delta\epsilon_{\mathrm{rec}}$ for the \textbf{a,} negative-mass ramp, \textbf{b,} positive-mass ramp and \textbf{c,} reference quench.
Circles show the experimental data; blue solid and orange dashed lines show full Bose--Hubbard evolution with $U=V$ and $(U-V)/h=4\,\mathrm{kHz}$, respectively.
For the ramps, $(J,V,\Delta)/h=(23.2,894,750)\,\mathrm{Hz}$ and $\delta/h$ changes linearly from zero to $\mp390\,\mathrm{Hz}$ over 100 ms.
For the quench, $(J,V,\Delta,\delta)/h=(76.2,821,600,-222.3)\,\mathrm{Hz}$.
Both calculations start from the ideal N\'eel state on $L=20$ sites and include the common Gaussian potential of Eq.~\eqref{eq:quench_trap}.
Time is measured from the first readout at $t_0=0.1\,\mathrm{ms}$.
The upper insets resolve the final 40\% of each protocol on the same enlarged vertical scale; shaded frames mark the magnified regions.
Error bars are pointwise 95\% intervals from 10,000 source-stratified acquisition-block bootstrap resamples, with the initial reference subtracted within each resample.}
\label{fig:unpaired_increment}
\end{figure}

Figure~\ref{fig:unpaired_increment} compares the negative- and positive-mass ramps with the reference quench.
The measured mean increments remain below eight percentage points throughout these protocols.
At the end of the negative and positive ramps, the increases are $4.0$ and $4.1$ percentage points, with 95\% intervals $[1.5,6.1]$ and $[1.3,6.8]$, respectively.
The quench gives an increase of $4.5$ percentage points, with a 95\% interval $[2.9,6.0]$.
These measurements demonstrate limited generation of additional pairing defects over the experimental evolution window.

To connect this measured increment to the atomic dynamics, we evolve the full Bose--Hubbard Hamiltonian from an ideal N\'eel state on an open chain.
The reconstructed occupations are obtained by taking the parity of each internal-state occupation, followed by the same pair assignment used for the images.
For example, an $\alpha\alpha\beta$ configuration gives one $\beta$ detection, which contributes to the unmatched count when it has no detected partner.
Local and nearest-neighbor projector expectation values give the mean detection and pair counts directly.
The numerical ratio is then formed from these mean counts, with the same initial-time subtraction as in Eq.~\eqref{eq:generated_defect_increment}.

We compare the $U=V$ Bose--Hubbard reference with a control in which $U$ is increased while $V$, $J$, $\Delta$, $\delta(t)$, the initial state and the trap remain fixed.
This changes the separation of same-species multiple occupations while preserving the projected parameters $\kappa$, $m$ and $g^2$.
It shifts the additional tunneling channels discussed in Sec.~\ref{sm:lgt:validity}, including the neighboring-doublon conversions $(2,2)\leftrightarrow(3,1)$ and $(2,2)\leftrightarrow(1,3)$, away from resonance.
At the experimental parameters, the calculated endpoint increments are $2.6$ percentage points for either ramp and $10.7$ percentage points for the quench; the measured quench increase is smaller than this ideal-initial-state reference.
Increasing $U$ reduces the corresponding numerical increments to $0.06$ and $1.1$ percentage points.
This control demonstrates that energetic separation of the additional atomic configurations suppresses defect production without changing the target gauge-theory parameters.

\FloatBarrier
\subsection{Benchmarking the hadron dynamics}
\label{sm:1d:benchmark}
\label{sm:1d:quench_target}

The hadron observables provide a direct comparison between the encoded gauge dynamics and the underlying atomic evolution.
We compare the experimental measurements with the target Hamiltonian and the full Bose--Hubbard model, using matched initial states, open boundaries and trapping potentials for the two theoretical descriptions.
The target calculation retains the complete reachable encoded sector, while the Bose--Hubbard calculation includes tunneling into additional atomic configurations.

\subsubsection{Mass-ramp observables}
\label{sm:1d:ramps}
\label{sm:1d:results}

\begin{figure}[htbp]
\centering
\includegraphics[width=\textwidth]{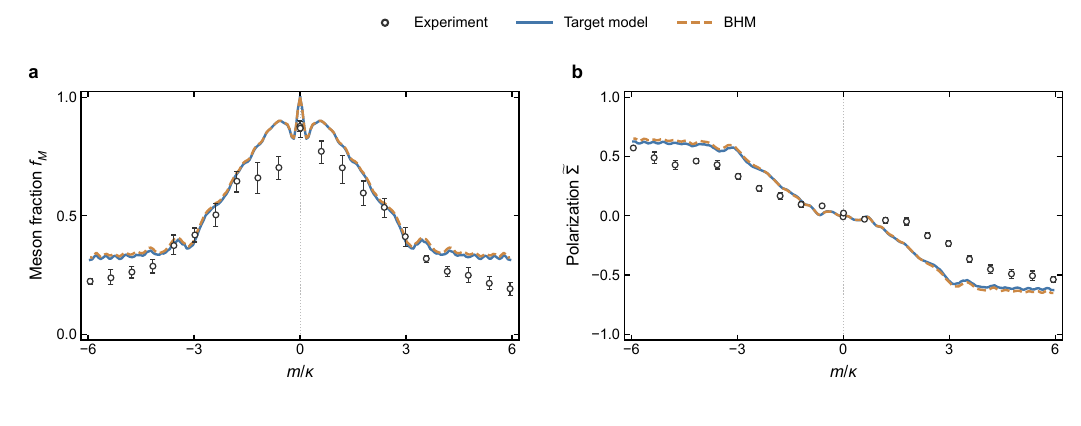}
\caption{\textbf{Atomic-model benchmark of the mass ramps.}
\textbf{a,} Meson-pair fraction $f_M$ and \textbf{b,} normalized staggered polarization $\widetilde\Sigma$ versus $m/\kappa$ for negative- and positive-mass ramps.
Circles denote experiment, blue solid lines the target Hamiltonian and orange dashed lines the full Bose--Hubbard model.
Both models use $L=20$, an ideal N\'eel initial state and the same Gaussian confinement.
The parameters are $J/h=23.2\,\mathrm{Hz}$, $U/h=V/h=894\,\mathrm{Hz}$ and $\Delta/h=750\,\mathrm{Hz}$, with $\delta/h$ ramped from zero to $\pm390\,\mathrm{Hz}$ over 100 ms.
These correspond to $g^2/\kappa\simeq11.7$ and endpoint masses $m/\kappa\simeq\pm5.94$.
Panel \textbf{b} uses all 1,379 nonempty strips from the 370 images at this coupling.
Error bars in \textbf{a} retain the 95\% intervals of the main-text data; those in \textbf{b} are 68.27\% intervals from 10,000 source-stratified acquisition-block resamples, keeping strips from the same image together.}
\label{fig:ramp_benchmark}
\end{figure}

Figure~\ref{fig:ramp_benchmark} compares the meson-pair fraction $f_M$ and normalized staggered polarization $\widetilde\Sigma$ for the two ramp directions.
Writing $D$ for the reconstructed meson count on both bond sublattices and $Q_{\mathrm p,a}$ for the reconstructed doublon counts, the model observables are
\begin{equation}
    f_M=\frac{\langle D\rangle}{\langle D+Q_{\mathrm p}+Q_{\mathrm a}\rangle},
    \qquad
    \widetilde\Sigma=
    \frac{\langle Q_{\mathrm p}-Q_{\mathrm a}\rangle}{\langle D+Q_{\mathrm p}+Q_{\mathrm a}\rangle}.
    \label{eq:benchmark_ramp_observables}
\end{equation}
The experimental meson fractions use the pooled counts of Fig.~\ref{fig:phase_mapping}c.
For the polarization, we display the means of the same five-chain-strip distributions shown in Fig.~\ref{fig:phase_mapping}d.

The two models give closely overlapping ramp trajectories, differing by at most $0.018$ in $f_M$ and $0.029$ in $\widetilde\Sigma$.
They capture meson depletion at large mass magnitudes and the polarization reversal between the baryonic and vacuum branches.
At the scan endpoints, the measured polarizations $(0.572,-0.538)$ compare with $\pm0.627$ for the target model and $\pm0.652$ for the Bose--Hubbard model.
The measured residual meson fractions are $0.223$ and $0.192$, compared with $0.314$ in the target model and $0.325$ in the Bose--Hubbard model.
Thus the full atomic model retains the target response of these slow-ramp observables.

\FloatBarrier
\subsubsection{Resolved hadron dynamics after a quench}
\label{sm:1d:quench_comparison}

\begin{figure}[htbp]
\centering
\includegraphics[width=\textwidth]{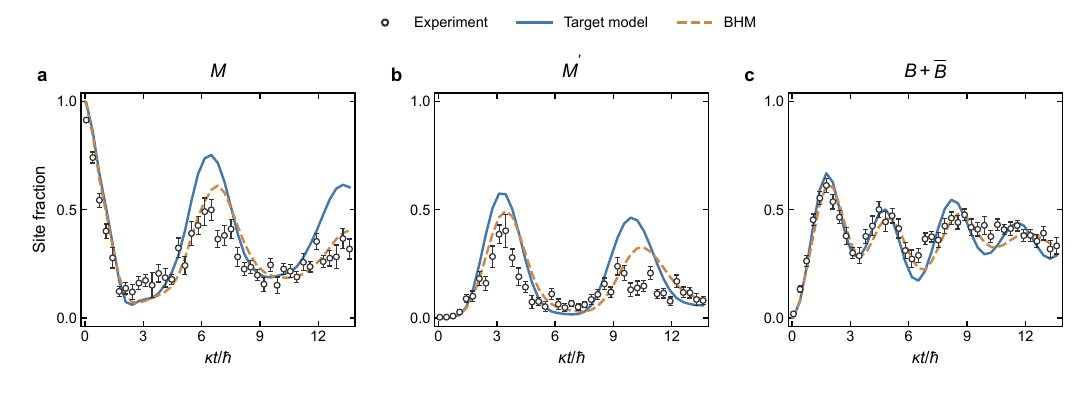}
\caption{\textbf{Resolved hadron populations in the target and atomic models.}
Site fractions of \textbf{a,} initial-bond mesons $M$, \textbf{b,} shifted-bond mesons $M'$ and \textbf{c,} baryons plus antibaryons $B+\bar B$ after the reference quench.
Circles and error bars reproduce the experimental means and bootstrap standard errors of Fig.~\ref{fig:mb_dynamics}b; blue solid and orange dashed lines show the target Hamiltonian and full Bose--Hubbard model.
The experimental ensemble contains 1,659 zero-net-baryon fragments over 41 times.
The models use $L=20$ and $(J,U,V,\Delta,\delta)/h=(76.2,821,821,600,-222.3)\,\mathrm{Hz}$, corresponding to $(m/\kappa,g^2/\kappa)\simeq(-1.03,5.47)$, with the same ideal N\'eel initial state and Gaussian potential.
The three populations use the site normalization of Eq.~\eqref{eq:benchmark_quench_observables}.}
\label{fig:quench_benchmark}
\end{figure}

The quench resolves how population moves between the two meson bond sublattices and the baryonic configurations (Fig.~\ref{fig:quench_benchmark}).
For the numerical chains we evaluate
\begin{equation}
    \rho_M^{\mathrm{initial}}=\frac{2\langle D_M\rangle}{L},\qquad
    \rho_M^{\mathrm{shifted}}=\frac{2\langle D_{M'}\rangle}{L},\qquad
    \rho_{B+\bar B}=\frac{\langle Q_{\mathrm p}\rangle+\langle H_{\mathrm a}\rangle}{L},
    \label{eq:benchmark_quench_observables}
\end{equation}
where $D_M$ and $D_{M'}$ count mesons on the initial and shifted bonds, respectively, and $H_{\mathrm a}$ counts detected holes on the antiparticle sublattice.
The two meson contributions sum to the total site fraction $\rho_M$ of Eq.~\eqref{eq:fig3_populations}.
For the full atomic model, baryon and antibaryon counts are evaluated separately from their reconstructed local configurations.
The site denominator is retained throughout the evolution, including when unpaired detections appear.
The experimental points reproduce the zero-net-baryon hadron populations of Fig.~\ref{fig:mb_dynamics}b.

Experiment and both models exhibit the same sequence: depletion of the initial mesons, growth of the baryonic population, and meson regeneration on the shifted bonds.
The Bose--Hubbard dynamics reduce the later revival amplitude of $M$ relative to the target calculation, bringing this population closer to the measurements.
Across the measured times, the unweighted root-mean-square deviations from experiment change from $0.150$ to $0.086$ for $M$ and from $0.133$ to $0.091$ for $M'$ when the target model is replaced by the full atomic model.
For $B+\bar B$, the corresponding deviations are $0.063$ and $0.053$.
The microscopic atomic dynamics therefore capture the reduced meson-revival amplitudes observed experimentally while preserving the target conversion sequence.

\FloatBarrier
\subsubsection{Density- and energy-dependent conversion}
\label{sm:1d:density_energy}

\begin{figure}[htbp]
\centering
\includegraphics[width=\textwidth]{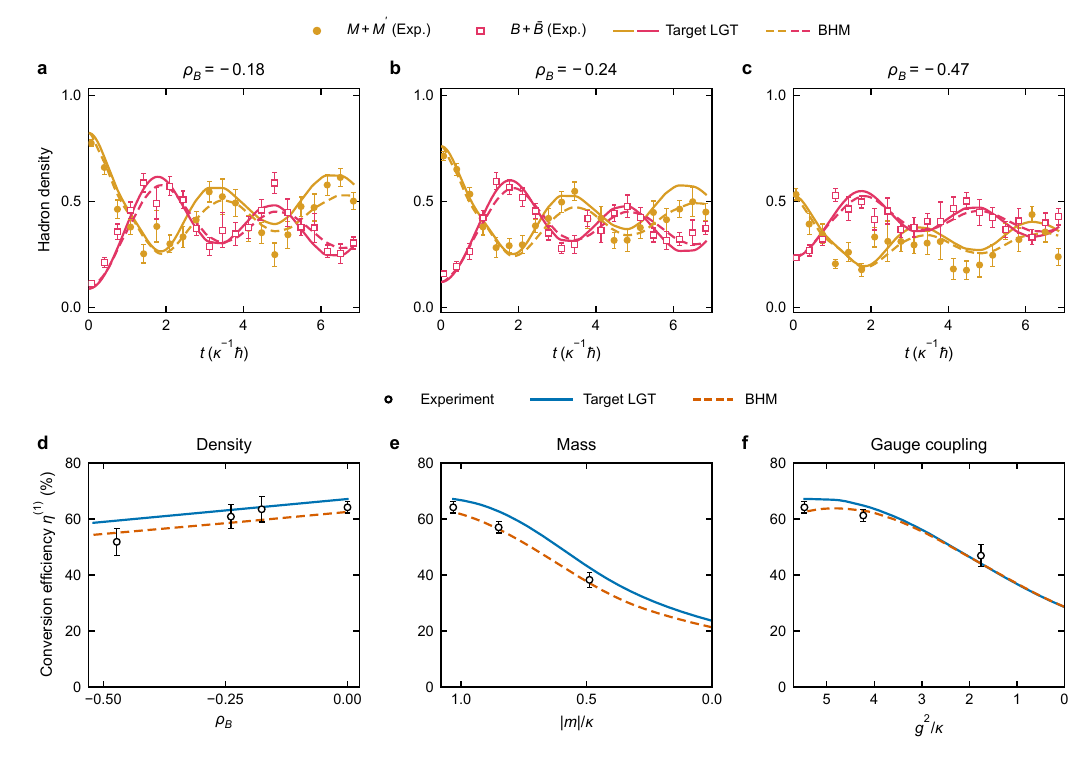}
\caption{\textbf{Density- and energy-dependent hadron dynamics in the target and atomic models.}
\textbf{a--c,} Total mesonic and baryonic site fractions for the three finite-density groups of Fig.~\ref{fig:hadron_response}b.
Gold circles show $M+M'$ and red squares show $B+\bar B$; solid and dashed curves show the target LGT and full Bose--Hubbard model, respectively.
\textbf{d--f,} First-maximum conversion responses versus net baryon density, mass and gauge coupling.
Open circles denote experiment, blue solid curves the target LGT and orange dashed curves the full atomic model.
Both models use open $L=20$ chains, the Table~\ref{tab:parameters} values and the common Gaussian potential of Eq.~\eqref{eq:quench_trap}.
Their finite-density ensembles use the same initial meson-removal configurations and weights.
Numerical responses use the full-chain population increments defined in Sec.~\ref{sm:1d:models}.
Experimental means and bootstrap standard errors reproduce Fig.~\ref{fig:hadron_response}.}
\label{fig:density_energy_benchmark}
\end{figure}

The comparison extends to the finite-density ensembles and energy scans of Fig.~\ref{fig:hadron_response} (Fig.~\ref{fig:density_energy_benchmark}).
For all three measured density groups, the target LGT and full Bose--Hubbard model reproduce the oscillatory exchange between mesonic and baryonic populations (Fig.~\ref{fig:density_energy_benchmark}a--c).
The full atomic calculation reduces the meson-revival amplitudes relative to the target model, bringing the total meson fractions closer to the experimental traces.

Both models also capture the principal density and energy dependence of the conversion response (Fig.~\ref{fig:density_energy_benchmark}d--f): conversion is suppressed at larger net baryon imbalance, smaller mass magnitudes and weak gauge coupling.
The response curves have similar shapes, with the full atomic model generally predicting lower efficiencies.
Together with the limited growth of the measured unpaired fraction and the ramp and reference-quench benchmarks, these comparisons validate meson--baryon conversion across the measured density and energy ranges.

\begingroup

The density dependence in Fig.~\ref{fig:density_energy_benchmark}d reflects changes in the local configurations through which mesons convert and reform.
In the initial-state ensembles used here, removing complete mesons introduces an antibaryon background, while each remaining meson retains the local pair-creation matrix element $\kappa/\sqrt2$.
A newly created baryon can then recombine with a neighbouring background antibaryon to form a meson on a shifted bond:
\begin{equation}
\ket{\bar B\,M}\;\longleftrightarrow\;
\ket{\bar B\,B\,\bar B}\;\longleftrightarrow\;
\ket{M'\,\bar B}.
\label{eq:background_recombination}
\end{equation}
This pathway transfers population back into mesonic configurations and reduces the growth of the total $B+\bar B$ population at the first maximum.

The local mechanism is particularly transparent in the homogeneous resonant limit, with the vacuum channel energetically separated as in Eq.~\eqref{eq:branch_separation}.
For an isolated three-state sequence in Eq.~\eqref{eq:background_recombination}, both couplings have magnitude $\kappa/\sqrt2$.
The antisymmetric superposition of the two mesonic configurations is uncoupled from the intermediate baryonic configuration.
Starting from $\ket{\bar B\,M}$, the probability of the intermediate state is therefore
\begin{equation}
P_{\bar B B\bar B}(t)=\frac12\sin^2\!\left(\frac{\kappa t}{\hbar}\right).
\label{eq:background_three_state}
\end{equation}
In a chain of neighbouring mesons, shifted-bond regeneration shares matter sites with adjacent conversion processes.
The occupation constraint thus couples the conversion of neighbouring mesons, producing a collective response that depends on their spatial arrangement.
Increasing the antibaryon background exposes more mesons to the local pathway in Eq.~\eqref{eq:background_recombination} and reduces the number of coupled neighbours.

Full Bose--Hubbard dynamics confirm this dependence on the initial arrangement.
For two mesons on a 12-site chain at the reference-quench parameters and the common Gaussian potential, the first-maximum conversion response is approximately $53\%$ when the mesons occupy adjacent central cells and $45\%$ when they occupy symmetrically separated cells.
Both configurations have the same net baryon density, $\rho_B=-2/3$.
The contrast also persists in a uniform potential, where their initial mean energies and energy variances coincide.
These results connect the density dependence in Fig.~\ref{fig:density_energy_benchmark}d to background-assisted meson regeneration and correlations between neighbouring conversion processes.

The density scan changes the configurations available for conversion, whereas the mass and gauge-coupling scans tune the relative energies of the baryonic and vacuum channels.
Equation~\eqref{eq:detunings} gives the corresponding detunings, $\varepsilon_B=3g^2/8+2m$ and $\varepsilon_V=3g^2/8-2m$.
The mass shifts these channels in opposite directions, setting their separation $\varepsilon_B-\varepsilon_V=4m$, whereas the gauge coupling changes their common electric-energy cost, $(\varepsilon_B+\varepsilon_V)/2=3g^2/8$.
The reference quench lies near $\varepsilon_B=0$, with the vacuum channel separated in energy.
Starting from this point on the negative-mass branch, reducing either $|m|$ or $g^2$ detunes baryon-pair creation and lowers the vacuum-channel detuning, accounting for the reduced conversion in Fig.~\ref{fig:density_energy_benchmark}e,f.

The two scans access different relations between the channels.
As $m$ approaches zero, the baryonic and vacuum configurations approach degeneracy, and meson population couples to both.
Changing $g^2$ at fixed $m$ instead preserves their energy separation while shifting both relative to the meson.
The full target and atomic calculations retain this competition together with meson regeneration on both bond sublattices.
The conversion response is thus governed by the available recombination pathways at finite density and their resonance conditions set by the mass and electric-field energy.
\endgroup

\FloatBarrier
\subsubsection{Numerical methods}
\label{sm:1d:models}
\label{sm:1d:ed}
\label{sm:1d:mps}
\label{sm:1d:validation}

The target Hamiltonian is evolved in its complete reachable sector on an open $L=20$ chain.
Both meson bond sublattices and the baryonic and vacuum conversion channels are retained.
For the mass ramps and reference quench, the initial state is a product of ten mesons on the initial bonds.
For the reference quench, tunneling is switched on instantaneously at the fixed final Hamiltonian.
For the ramp benchmarks, a $0.1\,\mathrm{ms}$ initial evolution at zero stagger precedes the 100-ms linear mass ramp in both models.

The spatial envelope enters as $\hat H_{\mathrm{trap}}=\sum_\ell\epsilon_\ell(\hat n_\ell^{\mathrm{at}}-1)$, where $\hat n_\ell^{\mathrm{at}}=\hat n_{\ell,\alpha}+\hat n_{\ell,\beta}$ is the total atomic occupation and both internal states experience the same potential.
We use a Gaussian profile
\begin{equation}
    \frac{\epsilon_\ell}{h}=A\left[1-\exp\!\left(-\frac{2x_\ell^2}{w^2}\right)\right],\qquad
    x_\ell=\left(\ell-\frac{L-1}{2}\right)a_s,\qquad
    A=\frac{Cw^2}{2a_s^2},
    \label{eq:quench_trap}
\end{equation}
with $a_s=0.630\,\mu\mathrm{m}$ and a $1/e^2$ radius $w=103\,\mu\mathrm{m}$, centered between the two middle sites.
The measured trap frequency $\nu_{\mathrm{eff}}=46\,\mathrm{Hz}$ fixes the curvature through $C=M_{\mathrm{Rb}}(2\pi\nu_{\mathrm{eff}})^2a_s^2/(2h)$, where $M_{\mathrm{Rb}}$ is the rubidium atomic mass.
This gives $C\simeq3.61\,\mathrm{Hz}$ and $A\simeq48.3\,\mathrm{kHz}$.
The same spatial potential is used for both Hamiltonians in Figs.~\ref{fig:unpaired_increment}--\ref{fig:density_energy_benchmark}.
It also enters the BHM curves in Figs.~\ref{fig:phase_mapping}d, \ref{fig:mb_dynamics} and \ref{fig:hadron_response}.

The target-state vectors are propagated using a Chebyshev expansion~\cite{talezer1984accurate}.
The ramp and reference-quench calculations use a $0.05\,\mathrm{ms}$ step, with the ramp Hamiltonian evaluated at each time-step midpoint.
The negative target ramp follows by reflection and exchange of the internal states, which preserve the initial state and symmetric trap and reverse the polarization.

The Bose--Hubbard calculations for Figs.~\ref{fig:phase_mapping}d, \ref{fig:mb_dynamics}, \ref{fig:hadron_response} and \ref{fig:unpaired_increment}--\ref{fig:density_energy_benchmark} retain the ten local states satisfying $n_{\ell,\alpha}+n_{\ell,\beta}\le3$.
This basis includes same-species doublons and all three-atom configurations.
We use matrix-product states with separately conserved $N_\alpha$ and $N_\beta$, evolved by second-order time-evolving block decimation in TeNPy~\cite{vidal2004efficient,Hauschild2018TeNPy}.
The experimental-parameter ramps and reference quench use maximum bond dimensions $\chi=256$ and $512$, respectively, and a $0.01\,\mathrm{ms}$ step.
The increased-$U$ controls use $\chi=128$ for the ramps and $256$ for the quench.
During the ramps, the Hamiltonian is updated at the midpoints of $0.1\,\mathrm{ms}$ intervals.
The finite-density and energy-scan calculations use $\chi=128$, increased to $256$ at the additional measured mass and gauge-coupling settings.

For readout observables, each internal-state occupation is mapped to its parity before mesons, doublons and holes are counted.
With $\hat n^{\mathrm{rec}}_\ell$ the detected occupation and $\hat P^{\mathrm{rec}}_{D,\ell}$ the detected $\alpha \beta$ doublon projector, the detected-hole projector is $\hat P^{\mathrm{rec}}_{0,\ell}=1-\hat n^{\mathrm{rec}}_\ell+\hat P^{\mathrm{rec}}_{D,\ell}$.
Its expectation value supplies the antibaryon count in Eq.~\eqref{eq:benchmark_quench_observables}.

The finite-density and energy-scan comparisons in Fig.~\ref{fig:density_energy_benchmark} use the same Table~\ref{tab:parameters} values and initial-state ensembles for both Hamiltonians.
Finite-density ensembles are formed by removing complete mesons from the ten-cell initial covering: all $2^{10}-1=1\,023$ nonempty configurations are evolved separately and assigned normalized weights proportional to $p_{\mathrm{ret}}^q(1-p_{\mathrm{ret}})^{10-q}$, where $q$ is the number of retained mesons and $\rho_B=q/10-1$.
The retention probability $p_{\mathrm{ret}}$ reproduces the measured mean density, and populations are averaged before baseline subtraction and normalization.

For either numerical model, we use the full-chain readout populations to define
\begin{equation*}
    \eta_{\mathrm{num}}(t)=
    \frac{\rho_{B+\bar B}^{\mathrm{num}}(t)-\rho_{B+\bar B}^{\mathrm{num}}(t_0)}
    {\rho_M^{\mathrm{num}}(t_0)},
    \qquad t_0=0.1\,\mathrm{ms}.
\end{equation*}
Here $\rho_{B+\bar B}^{\mathrm{num}}=\langle N_B+N_{\bar B}\rangle/L$, and $\rho_M^{\mathrm{num}}$ includes both meson bond sublattices.
Subtracting the initial population removes the baryonic background present at $t_0$.
For the target LGT, the excess baryonic population is conserved, so its initial-subtracted response equals that of the paired population.
Experimental efficiencies use the selected-fragment estimator in Eq.~\eqref{eq:fig4_balanced_population}.
The numerical conversion responses are evaluated at the first maximum, located by cubic-spline interpolation within $1.1$--$3.6\,\mathrm{ms}$.

\section{Data processing and statistical analysis}
\label{sm:analysis}

\subsection{Experimental data and sample selection}
\label{sm:analysis:data}
\label{sm:analysis:reconstruction}
\label{sm:analysis:ensembles}
\label{sm:analysis:robustness}

Spin-resolved occupations are reconstructed from the fluorescence images using the Stern--Gerlach readout described in Sec.~\ref{sm:setup:readout}.
We identify directed neighboring $(\alpha,\beta)$ pairs as mesons and on-site $\alpha\beta$ pairs as doublons; the particle and antiparticle sublattices distinguish baryonic from vacuum configurations according to Sec.~\ref{sm:lgt:observables}.
Each image records an independently prepared experimental realization.

For the mass ramps in Fig.~\ref{fig:phase_mapping}c--e, we use the full reconstructed field of view of each accepted image, without selecting chains by length or net baryon number.
The dataset contains 1,565 images across 88 conditions.
Writing $D_s$ for the number of mesons on both bond sublattices in image $s$, and $Q_{\mathrm p,s}$ and $Q_{\mathrm a,s}$ for the doublon counts on the particle and antiparticle sublattices, we obtain
\begin{equation}
    f_M=\frac{\sum_sD_s}{\sum_s(D_s+Q_{\mathrm p,s}+Q_{\mathrm a,s})}.
    \label{eq:fig2_pair_fraction}
\end{equation}
Counts are pooled before taking the ratio; unmatched single atoms are omitted from the pair count while their parent images remain in the ensemble.
For the polarization distributions in Fig.~\ref{fig:phase_mapping}d, each image is divided into four strips of five neighboring chains.
The 370 images at this coupling yield 1,379 nonempty strips, each contributing equal weight with
\begin{equation}
    \widetilde\Sigma=\frac{Q_{\mathrm p}-Q_{\mathrm a}}{D+Q_{\mathrm p}+Q_{\mathrm a}},
    \label{eq:fractions}
\end{equation}
where the counts refer to the strip.
Histogram error bars are 68.27\% bootstrap percentile intervals obtained by resampling images within each source, keeping all strips from the same image together.
The representative averages in Fig.~\ref{fig:phase_mapping}a,b use separate display ensembles selected by cloud shape and local uniformity: 45 mesonic, 50 baryonic and 42 vacuum images, pooled from the corresponding regimes.

\label{sm:analysis:quench}
\label{sm:fig3_analysis}
For the quench data in Figs.~\ref{fig:mb_dynamics} and \ref{fig:hadron_response}, we select encoded fragments from the reconstructed chains.
We first exclude parent chains touching the reconstruction boundary, and then segment the remaining chains at unmatched single atoms, retaining complete mesons, doublons and holes.
Nonempty fragments with at least eight sites enter the analysis.
The role of unmatched atoms as fragment boundaries, including their behavior in the full atomic model, is analyzed in Sec.~\ref{sm:1d:boundaries}.
Figure~\ref{fig:mb_dynamics} uses the zero-net-baryon subset, $N_B=N_{\bar B}$: 1,659 fragments at 41 times between 0.1 and 20.1 ms, drawn from the reference dataset of 2,233 valid images.
The population curves use fragments across the full field of view, without the circular display window used for the local maps.

\begin{figure}[htbp]
\centering
\includegraphics[width=\textwidth]{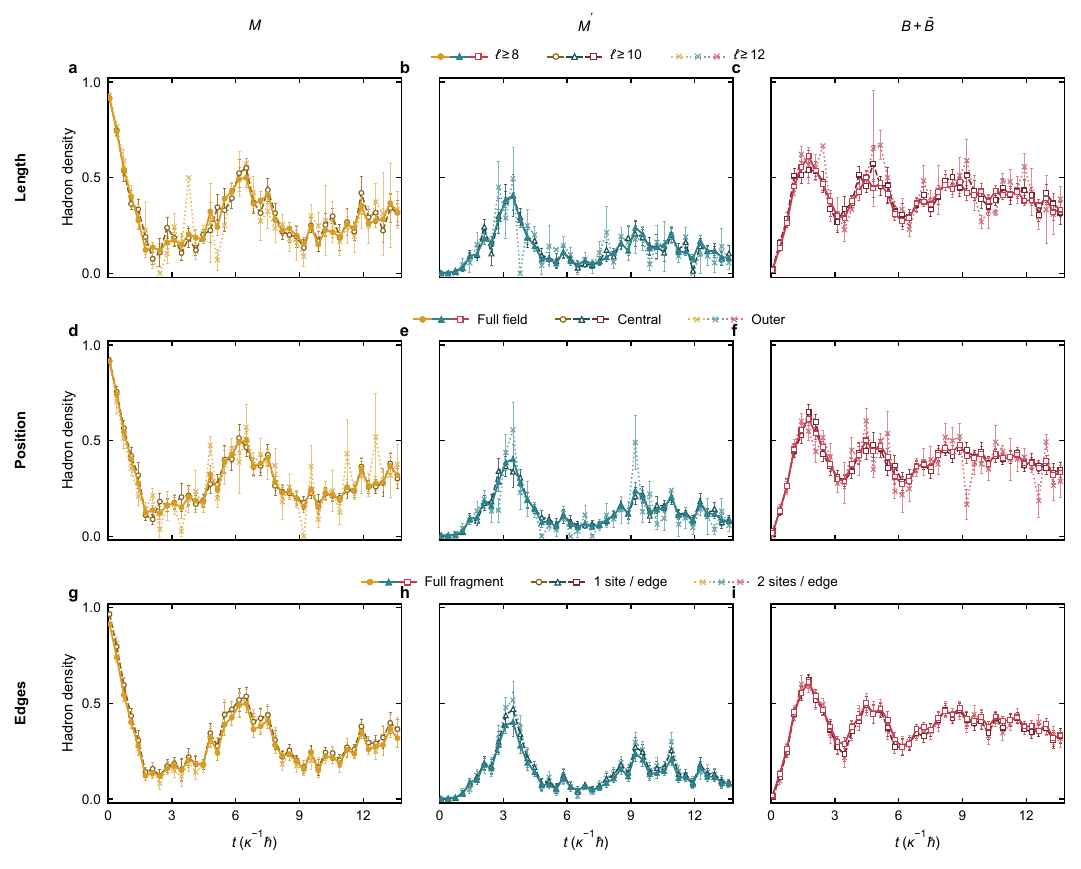}
\caption{\textbf{Robustness of the quench dynamics to fragment selection and edge treatment.}
Columns show the site fractions of initial-bond mesons $M$ (gold), shifted-bond mesons $M'$ (teal), and baryons plus antibaryons $B+\bar B$ (red), matching Fig.~\ref{fig:mb_dynamics}.
\textbf{a--c,} Minimum fragment lengths $\ell\geq8$, 10 and 12, containing 1,659, 670 and 251 fragments, respectively.
\textbf{d--f,} Full-field, central and outer subsets.
The central subset satisfies $|r_c-20|\leq8$, where $r_c$ is the fragment midpoint in the registered lattice coordinates; the outer subset is its complement.
These subsets contain 1,231 and 428 fragments, respectively.
\textbf{g--i,} Full fragments and interior windows obtained by excluding one or two sites at each end.
The same original zero-net-baryon fragments are retained in the edge comparison, with site fractions normalized to the remaining window length.
Mesons are identified before cropping, and each retained mesonic site contributes to its corresponding population.
Solid, dashed and dotted lines distinguish the three choices in each row and connect experimental means.
Error bars denote bootstrap standard errors from 10,000 resamples of whole time scans within each data source.
Points with a single contributing image are shown without an error bar.}
\label{fig:quench_selection_robustness}
\end{figure}

We test this fragment-based population analysis against changes in fragment length, position and the treatment of edge sites (Fig.~\ref{fig:quench_selection_robustness}).
These choices preserve the characteristic sequence of initial-meson depletion, baryon--antibaryon production and meson regeneration on shifted bonds.
The stricter length and outer-region selections contain fewer fragments and show larger point-to-point scatter.
Excluding one or two sites at each edge changes the population traces modestly and retains a pronounced $M'$ revival in the fragment interior.
These comparisons support the robustness of the observed hadronic dynamics to the tested fragment selections and observation windows.

For fragment $f$ of length $\ell_f$, the mesonic and baryonic site fractions are averaged with equal fragment weights:
\begin{equation}
    \rho_M(t)=\frac{1}{N_t}\sum_{f\in t}\frac{2N_{M,f}}{\ell_f},
    \qquad
    \rho_{B+\bar B}(t)=\frac{1}{N_t}\sum_{f\in t}\frac{N_{B,f}+N_{\bar B,f}}{\ell_f},
    \label{eq:fig3_populations}
\end{equation}
where $N_t$ is the number of selected fragments at time $t$ and $N_{M,f}$ includes both meson bond sublattices.
Resolving this count by bond parity gives the separate $M$ and $M'$ populations in Fig.~\ref{fig:mb_dynamics}b; their sum gives the total meson fraction used in Fig.~\ref{fig:hadron_response}.
The remaining sites can occupy the encoded vacuum configurations.
Population error bars denote bootstrap standard errors.

\label{sm:analysis:conversion}
\label{sm:fig4_analysis}
For Fig.~\ref{fig:hadron_response}b, the same fragments are grouped by their measured net baryon density, $\rho_{B,f}=2(N_{B,f}-N_{\bar B,f})/\ell_f$.
We report the density groups $\rho_B=0$, $[-0.2,0)$, $[-0.25,-0.2)$ and $[-0.5,-0.4)$, for which the available samples support stable estimation of the first conversion maximum.
The nonzero-density coordinates are the equal-time means of the fragment-averaged densities over 0.1--10.1 ms.
To measure conversion after subtracting the excess baryonic population associated with the net imbalance, we use
\begin{equation}
    p(t)=\frac{1}{N_t}\sum_{f\in t}\frac{2\min(N_{B,f},N_{\bar B,f})}{\ell_f},
    \qquad
    \eta(t)=\frac{p(t)-p(t_0)}{\rho_M(t_0)},
    \label{eq:fig4_balanced_population}
\end{equation}
with the initial reference fixed at the earliest readout, $t_0=0.1\,\mathrm{ms}$.
The mass and gauge-coupling scans in Fig.~\ref{fig:hadron_response}c use the zero-net-baryon group.

\label{sm:analysis:visualization}
Local maps are formed by dividing the count of each configuration by the number of valid selected observations at the corresponding site or bond origin.
For the circular maps in Fig.~\ref{fig:mb_dynamics}b, counts and observation numbers are pooled separately over a $3\times3$ neighborhood of equal-parity positions before taking their ratio.
For Fig.~\ref{fig:mb_dynamics}c, the same quantities are pooled across parallel chains at fixed registered coordinates and over three equal-parity positions along the chain, retaining links whose endpoints belong to the same fragment.
The link Casimir is assigned as $E^2=0$ inside a meson and $E^2=3/4$ on other valid links; the matter excitation density follows the particle--hole dictionary in Sec.~\ref{sm:lgt:observables}.
These maps use the measured time points without temporal interpolation.
The nonlinear color normalizations of Fig.~\ref{fig:phase_mapping}b and the matter-density map in Fig.~\ref{fig:mb_dynamics}c act on the display values, leaving the population statistics unchanged.

\FloatBarrier
\subsection{Fitting procedures}
\label{sm:analysis:fits}
\label{sm:analysis:ramp}
\label{sm:fig2_analysis}

\textit{Mass-ramp fits.}
The positive- and negative-mass ramps at each gauge coupling are fitted independently with an empirical sigmoid that is flat at the origin:
\begin{equation}
    f(x)=b+(H_0-b)\frac{z(x)}{q+(1-q)z(x)},
    \qquad
    z(x)=\exp\!\left[-\frac{\sqrt{x^2+s^2}-s}{w}\right],
    \quad x=\frac{|m|}{\kappa}.
    \label{eq:fig2_empirical_ramp_fit}
\end{equation}
The smoothing scale $s$ is fixed at half the mass-sampling interval, giving $s\simeq0.30$.
The four parameters are optimized separately for each ramp by weighted least squares, with bounds $0\le b\le0.49$, $0.5\le H_0\le1$, $10^{-6}\le q\le1$ and $w$ restricted to approximately $0.035$--$14.1$.
The weights are the inverse variances of the measured meson fractions.
For each fitted curve, the conversion mass is the position at which $f(x)=[f(0)+f(x_{\mathrm{end}})]/2$, with $x_{\mathrm{end}}\simeq5.94$.
Thus each ramp is evaluated at half of its fitted decrease over the measured scan, with the sign of the mass restored for the corresponding branch.

\label{sm:analysis:statistics}
The statistical distribution of $f_M$ is obtained from 20,000 source-stratified bootstrap resamples: complete acquisition loops are resampled within each scanned source, and individual images within the additional endpoint source.
A subset of 1,000 resampled curves is refitted, and the finite-scan midpoint is evaluated for each fit to propagate the uncertainty to the conversion masses.
Error bars in Fig.~\ref{fig:phase_mapping}c,e are the resulting 95\% percentile intervals.
For Fig.~\ref{fig:phase_mapping}e, each branch is then fitted as a linear relation between $m/\kappa$ and $g^2/\kappa$, weighted by the inverse bootstrap variance of the conversion masses.
The fitted relations are algebraically inverted for the displayed $g^2(m)$ annotations; dashed segments extend these fits outside the measured coupling range.

\textit{First-peak fits.}
For all density, mass and gauge-coupling conditions in Fig.~\ref{fig:hadron_response}, we fit $p(t)$ over the same eight measurement times from 0.1 to 3.6 ms using
\begin{equation}
    p_{\mathrm{fit}}(t)=b+(1-b)a\cos^2\!\left[\frac{\pi(t-t_p)}{T}\right],
    \qquad
    \eta^{(1)}=\frac{b+(1-b)a-p(t_0)}{\rho_M(t_0)}.
    \label{eq:fig4_peak_estimator}
\end{equation}
Here $t_p$ locates the peak and $T$ controls the shape of the fitted first lobe.
The bounds are $0\le b\le0.5$, $0\le a\le1$, $1.1\le t_p\le3.6\,\mathrm{ms}$ and $3.6\le T\le12\,\mathrm{ms}$.
Weighted least squares uses fixed weights derived from the fragment standard errors, with an error floor equal to the larger of $10^{-4}$ and one quarter of the median nonzero standard error for that condition; multiple initial guesses are used.
The upper curves in Fig.~\ref{fig:hadron_response}c use these fits, expressed as $\eta(t)$.

For the quench measurements, we generate 10,000 bootstrap resamples within each data source.
Each repeated time scan is sampled with replacement as a whole, keeping its recorded images and selected fragments together.
The hadronic populations and density groups are recomputed jointly in each resample, preserving the equal-fragment averages defined above.
For $\eta(t)$ and $\eta^{(1)}$, the initial reference and mesonic normalization are recalculated in every resample; the first peak is refitted with the same weights, parameter bounds and time window.
The shared zero-density reference is reused across the density, mass and gauge-coupling comparisons.
Reported standard errors are the sample standard deviations of the bootstrap estimates, evaluated on resamples with nonempty contributing groups.
The first-peak uncertainties use 9,955 valid estimates for $\rho_B\in[-0.5,-0.4)$, 9,983 for $m/\kappa\simeq-0.49$, and 10,000 for the other conditions.
The BHM population and conversion-response curves use the numerical procedures described in Sec.~\ref{sm:1d:quench_target}.

\end{document}